\documentclass[preprint,12pt]{elsarticle}

\usepackage{graphicx}  
\usepackage{microtype} 
\usepackage{amsmath} 
\usepackage{amssymb}

\usepackage[colorlinks=false,hidelinks]{hyperref}
\usepackage{color}
\usepackage{epstopdf}

\usepackage{natbib}

\newcommand{\beq}{\begin{equation}}  
\newcommand{\eeq}{\end{equation}}    
\newcommand{\beqa}{\begin{eqnarray}} 
\newcommand{\eeqa}{\end{eqnarray}}   

\biboptions{semicolon,square}

\newcounter{bla}

\def\ie{{\it i.e.\/}}
\def\eg{{\it e.g.\/}}

\def\1o2{\textstyle {\frac{1}{2}}}

\def\betab{\mbox{\boldmath $\beta$}}
\def\d{{\rm d}}

\def\dotprod{\!\cdot\!}

\def\req#1{(\ref{#1})}

\journal{Nuclear Instruments and Methods B}

\begin{document}

\begin{frontmatter}


\title{Nuclear effects in proton transport and dose calculations}

\author[a]{Francesc Salvat\corref{author}}
\author[b]{Jos\'{e} Manuel Quesada}

\cortext[author] {Corresponding author.\\\textit{E-mail address:}
francesc.salvat@ub.edu}

\address[a]{
Facultat de F\'{\i}sica (FQA and ICC), Universitat de Barcelona,
Diagonal 645, 08028 Barcelona, Spain
}

\address[b]{
Departamento de F\'{\i}sica At\'{o}mica, Molecular y
Nuclear, Universidad de Sevilla, 18071 Sevilla, Spain
}

\begin{abstract}

\noindent Interactions of protons with nuclei are modeled in a form that
is suitable for Monte Carlo simulation of proton transport. The
differential
cross section (DCS) for elastic collisions of protons with neutral atoms
is expressed as the product of the Rutherford DCS, which describes
scattering by a bare point nucleus, and two correction factors that
account for the screening of the nuclear charge by the atomic electrons
and for the effect of the structure of the nucleus. The screening
correction is obtained by considering the scattering of the projectile
by an atom with a point nucleus and the atomic electron cloud described
by a parameterization of the Dirac-Hartree-Fock-Slater self-consistent
electron density. The DCS for scattering by this point nucleus atom is
calculated by means of the eikonal approximation. The nuclear correction
to the DCS for elastic collisions is calculated by conventional
partial-wave analysis with a global optical-model potential that
describes the interaction with the bare nucleus. Inelastic interactions
of the projectile with target nuclei are described by using information
from data files in ENDF-6 format, which provide cross sections,
multiplicities, and angle-energy distributions of all reaction products:
light ejectiles (neutrons, protons, \ldots), gammas, as well as
recoiling heavy residuals. These interaction models and data have been used
in the Monte Carlo transport code {\sc penh}, an extension of the
electron-gamma code {\sc penelope}, which originally accounted for
electromagnetic interactions only. The combined code system {\sc
penh}/{\sc penelope} performs simulations of coupled
electron-photon-proton transport. A few examples of simulation results
are presented to reveal the influence of nuclear interactions on proton
transport processes and on the calculation of dose distributions from
proton beams.

\end{abstract}


\end{frontmatter}

\section{Introduction \label{sec1}}

Monte Carlo simulation is the prime tool for solving radiation transport
problems, and finds applications in fields as diverse as microscopy, high-energy physics, radiation
metrology and dosimetry, medical physics, and a variety of
spectroscopic techniques. Simulations of energetic
charged particles are difficult because of the large number of
interactions undergone by these particles before being brought
to rest. General-purpose Monte Carlo codes for high-energy radiation
transport (\eg,
{\sc etran} \cite{BergerSeltzer1988a,BergerSeltzer1988b, BergerSeltzer1988c},
{\sc its3} \cite{Halbleib1992},
{\sc egs4} \cite{Nelson1985},
{\sc geant3} \cite{Brun1987},
{\sc egs}nrc \cite{KawrakowRogers2001},
{\sc mcnp} \cite{MCNP03},
{\sc geant4} \cite{Agostinelli2003, Allison2006, Allison2016},
{\sc fluka} \cite{Ferrari2005},
{\sc egs5} \cite{Hirayama2006}
{\sc mcnp6} \cite{Goorley2013})
utilize detailed event-by-event simulation for photons, while charged
particles are simulated by means of a combination of class I and class
II schemes (see Ref.\ \cite{Berger1963}). In class I or condensed
simulation schemes, the trajectory of a charged particle is split into
segments of predefined length and the cumulative energy loss and angular
deflection resulting from the interactions along each segment are
sampled from approximate multiple scattering theories. The fluctuation
of the energy loss along each segment, is usually accounted for by means
of the energy-straggling theories of Landau \cite{Landau1944}, Blunk and
Leisegang \cite{BlunckLeisegang1950}, or modifications of these
theories
\cite{BichselSaxon1975}.  Typically, the accumulated angular deflection
along a segment is described by means of the multiple-scattering
theories of Moli\`{e}re \cite{Moliere1948}, Goudsmit and Saunderson
\cite{GoudsmitSaunderson1940, GoudsmitSaunderson1940b}, or Lewis
\cite{Lewis1950}. As these theories provide only a partial description
of the transport process, they need to be complemented with some sort of
approximation to determine the net spatial displacement of the
transported particle at the end of the segment, the particle-step
transport algorithm \cite{BielajewRogers1987}.

Class II, or mixed, schemes simulate individual hard interactions (\ie,
interactions with energy loss $W$ or polar angular deflection $\theta$
larger than certain cut-offs $W_{\rm c}$ and $\theta_{\rm c}$) from
their
restricted differential cross sections (DCSs), and the effect of the
soft interactions (with $W$ or $\theta$ less than the corresponding
cut-off) between each pair of hard interactions is described by means of
multiple-scattering approximations. Class II schemes not only describe
the hard interactions accurately (\ie, according to the adopted DCSs)
but also simplify the description of soft events, which have a lesser
impact on the transport process. Indeed, by virtue of the central limit
theorem, the global effect of a succession of interactions involving
only small angular deflections and small energy transfers is nearly
Gaussian \cite{RossiGreisen1941,Eyges1948, Salvat2015} and it can be
described in terms of a few integrals of the DCS restricted to soft
interactions.

To the best of our knowledge, {\sc penelope} \cite{Baro1995,Salvat2015}
is the only general-purpose code that consistently uses class II
simulation for all interaction processes of electrons and
positrons,
while other codes generally use a combination of class I, class II and
detailed schemes. In
addition, {\sc penelope} implements a simple tracking scheme (the
so-called random-hinge method) that describes spatial displacements
fairly accurately. An additional advantage of a class II code is that,
when the cut-offs $W_{\rm c}$ and $\theta_{\rm c}$ are set to zero, it
performs detailed (event by event) simulations. Since detailed
simulation provides an exact reproduction of the transport process
conducted by the adopted DCSs (apart from statistical uncertainties),
the accuracy of class II simulation can be readily verified by comparing
results from detailed and class II simulations of the same arrangement.

The {\sc penelope} tracking algorithm was also implemented in the proton
transport code {\sc penh} \cite{Salvat2013} which originally only
considered electromagnetic interactions, with atomic nuclei represented
as point charges. In the present article we describe physics models and
sampling methods that were developed to account for nuclear effects in
{\sc penh}, or in any other class II proton simulation code. We focus on
two dominant features that can be included in a class II transport
simulation without altering the structure of the code significantly. On
the one hand, protons with kinetic energy higher than a few MeV are able
to overcome the Coulomb barrier of the nucleus and induce nuclear
reactions, which act as a sink for the protons and as a source of
energetic
reaction products. On the other hand, the finite size and the structure
of the nucleus have an effect on the elastic collisions of protons,
which alter the DCS at intermediate and large angles by introducing an
oscillatory structure.

A detailed microscopic description of the proton-nucleus interaction is
not practicable because of the complexity of quantum mechanical and
semiclassical treatments. In addition, theoretical calculations of
nuclear reactions provide only partial information on the process and in
limited energy ranges. Looking for the widest flexibility of the code,
we simulate proton-induced nuclear reactions by using information from
nuclear databases in the standard ENDF-6 format \citep{ENDF2018}. These
databases provide the reaction cross section $\sigma_{\rm nr}(E)$ and a
statistical description of the products emitted in a reaction, as
functions of the kinetic energy $E$ of the proton, the average number of
released products (prompt gammas, neutrons, protons, deuterons, tritons,
$^3$He, alphas, and residual nuclei), and their angular and energy
distributions.

The usual practice in electromagnetic transport simulation codes is to
describe elastic collisions by an atomic DCS calculated by considering
the target atom as a frozen distribution of electronic charge that
screens the Coulomb field of the (point) nucleus. However, the DCS for
elastic collisions of protons with atoms is also sensitive to nuclear
effects. Although the ENDF-6 files do account for the effect of the
finite size and the structure of the nucleus on the elastic DCS, which
is described as a correction to the Rutherford DCS, they disregard
screening effects. The DCSs adopted in the present study were computed
by combining accurate calculations of scattering by the point nucleus
screened by the atomic electron cloud and by the bare finite nucleus (by
using the same methodology as in the generation of the ENDF-6 files,
\ie, conventional partial-wave analysis with a realistic optical-model
potential). We have calculated a numerical database of DCSs for elastic
scattering of protons by atoms (averaged over naturally occurring
isotopes) of the elements with atomic number $Z$ from 1 to 99 and proton
kinetic energies between 100 keV and 1 GeV. The proton transport code
{\sc penh} has been modified to account for nuclear effects. The new
code simulates elastic collisions by using the numerical DCSs in our
database, and it describes nuclear reactions according to the
information provided by nuclear databases in the generic ENDF-6 format.

The present article is organized as follows. Section 2 describes the
theory and calculations of DCSs for elastic collisions of protons with
neutral atoms together with the sampling strategy adopted in the
simulation code. An overview of the
simulation of inelastic electronic excitations by proton
impact, which follows the same scheme as in the previous version of
{\sc penh} \cite{Salvat2013}, is given
in Section 3. Section 4 gives a detailed description
of the simulation of nuclear reactions based on information from ENDF-6
formatted files. Sample simulation results from {\sc penh} are presented
in Section 5, where the influence of nuclear effects on the dose
distributions of proton beams in water is discussed. Finally, we
offer our concluding comments in Section 6.


\section{Elastic collisions\label{sec2}}

Let us consider elastic collisions of protons (mass $m_{\rm p}$ and charge
$+e$) with atoms of the element of atomic number $Z$. In order to cover
the range
of proton energies of interest in proton therapy, up to about 300 MeV,
we shall use relativistic collision kinematics. The simulation code
transports particles in the laboratory (L) frame, where the target atom
is at rest and the proton moves with kinetic energy $E$ before the
collision. For simplicity, we consider that the $z$ axis of the
reference frame is parallel to the linear momentum of the proton, which
is given by
\beq
{\bf p} = c^{-1} \sqrt{E (E + 2 m_{\rm p} c^2})
\, \hat{\bf z},
\label{n.1}\eeq
where $c$ is the speed of light in vacuum. The total energy of the
projectile proton is
\beq
{\cal W} = E + m_{\rm p} c^2 = \sqrt{m_{\rm p}^2 c^4 + c^2 p^2}.
\label{n.2}\eeq
We recall the general relations
\beq
p = \beta \gamma \, m_{\rm p} c \qquad \mbox{and} \qquad
E=(\gamma-1) m_{\rm p} c^2
\label{n.3}\eeq
where
\beq
\beta= \frac{v}{c} = \frac{\sqrt{E(E+2m_{\rm p} c^2)}}{E+m_{\rm p}c^2}
\label{n.4}\eeq
is the speed of the proton in units of $c$ and
\beq
\gamma = \sqrt{\frac{1}{1-\beta^2}} = \frac{E+m_{\rm p}c^2}{m_{\rm
p}c^2}
\label{n.5}\eeq
is the total energy in units of the proton rest energy.

Elastic collisions involve a certain transfer of kinetic
energy to the target atom, which is easily accounted for by sampling the
collisions in the center-of-mass (CM) frame, which moves with velocity
\beq
{\bf v}_{\rm CM} = \betab_{\rm CM} c = \frac{c^2 {\bf p}}{E +
m_{\rm p} c^2 + M_{\rm A} c^2},
\label{n.6}\eeq
where $M_{\rm A}$ is the mass of the atom. In the CM frame the linear
momenta of the proton and the atom before the collision are,
respectively, ${\bf p}'_{\rm pi} = {\bf p}'_0$ and ${\bf p}'_{\rm Ai} =
-{\bf p}'_0$, with
\beq
{\bf p}'_0 = \frac{M_{\rm A} c^2}{\sqrt{(m_{\rm p} c^2
+ M_{\rm A} c^2)^2 + 2 M_{\rm A} c^2 \; E}} \; {\bf p}.
\label{n.7}\eeq
Quantities in the CM frame are denoted by primes. After the elastic
collision, in CM the projectile moves with momentum $p'_{\rm
pf}=p'_0$ in a direction defined by the polar scattering angle $\theta$ and
the azimuthal scattering angle $\phi$, and the target atom recoils with
equal momentum $p'_{\rm Af}=p'_0$ in the opposite direction. The
final energies and directions of the proton and the atom in the L frame
are obtained by means of a Lorentz boost with velocity $-{\bf v}_{\rm
CM}$. Thus, elastic collisions are completely determined by the DCS $\d
\sigma/\d \Omega'$ in the CM frame.


\subsection{Interaction potential}

In calculations of elastic collisions, the interaction between the
projectile proton and the target atom is described by a
central potential. The central-potential approximation
allows the calculation of the DCS by the conventional partial-wave
expansion method, which provides a nominally exact solution of the
scattering wave equation. As a first approximation, the atomic nucleus
can be considered as a point particle, in which case the interaction
potential is
\beq
V_{\rm scr}(r) = \frac{Ze^2}{r} \,\Phi(r)
\label{n.8}\eeq
where $r$ is the distance between the proton and the nucleus and
$\Phi(r)$ is the screening function, which accounts for the shielding of
the nuclear charge by the atomic electrons. To facilitate calculations,
we use approximate screening functions having the analytical form
\beq
\Phi(r) = \sum_{i=1}^3 A_i \exp(- a_i r)
\quad \mbox{with} \quad \sum_{i=1}^n A_i = 1,
\label{n.9}\eeq
with the parameters given by \cite{Salvat1987}, which were determined by
fitting the self-consistent Dirac-Hartree-Fock-Slater (DHFS) atomic
potential of neutral free atoms. The advantage of using this
representation of the potential is that a good part of the calculations
can be performed analytically. The multiple scattering theory of
Moli\`{e}re \cite{Moliere1947, Moliere1948} is based on a
parameterization of the DCS calculated with the screening function of
Thomas-Fermi atoms approximated in the form \req{n.9}. Numerical
relativistic partial-wave calculations \cite{SalvatFernandezVarea2019}
of elastic collisions of electrons and positrons (mass $m_{\rm e}$ and
charge $\pm e$) show that the DHFS potential and the
analytical approximation \req{n.9} yield practically equivalent DCSs for
projectiles with kinetic energies higher than about 1 keV.

The interaction of the projectile proton with a bare nucleus of the
isotope $^AZ$ having atomic number $Z$ and mass number $A$ can be
described by a phenomenological complex optical-model potential
\beq
V_{\rm nuc} (r) = V_{\rm opt}(r) + {\rm i} W_{\rm opt}(r),
\label{n.10}\eeq
where the first term is a real potential that reduces to the Coulomb
potential at large radii and the second term, ${\rm i} W_{\rm nuc}(r)$,
is an absorptive (negative) imaginary potential which accounts for the
loss of protons from the elastic channel caused by inelastic processes.
Parameterizations of optical-model potentials have been proposed by
various authors (see, \eg, Refs.\ \cite{BecchettiGreenlees1969,
Hodgson1971, KoningDelaroche2003}). They are generally written as a
combination of Woods--Saxon volume terms,
\begin{subequations}
\label{n.11}
\beq
f(R,a;r) = \frac{1}{1+\exp[(r-R)/a]}\, ,
\label{n.11a}\eeq
and surface derivative (d) terms,
\beqa
g(R,a; r) &=& \frac{\d}{\d r} \, f(R,a;r)
\nonumber \\ [2mm]
&=& \frac{1}{a}\, f(R,a;r) \left[ f(R,a;r) - 1 \right].
\label{n.11b}\eeqa
\end{subequations}
The parameters in these functions are the radius $R$ and the diffuseness
$a$; typically, the radius is expressed as $R=r_0 A^{1/3}$. We consider
a generic potential having the form
\beqa
V_{\rm nuc} (r)
&=& V_{\rm v} (E;r) + V_{\rm d} (E;r) + V_{\rm c} (r) + V_{\rm so}
(E;r) \, 2 \, {\bf L} \dotprod {\bf S}
\nonumber \\ [2mm]
&& \mbox{}
+ {\rm i} \left[ W_{\rm v} (E;r) + W_{\rm d} (E;r)
+ W_{\rm so} (E;r) \, 2 \, {\bf L} \dotprod {\bf S}
\rule{0mm}{3.5mm}\right]
\label{n.12}\eeqa
with the following terms: \\ [2mm]
\noindent 1) Real volume potential:
\begin{subequations}
\label{n.13}
\beq
V_{\rm v} (E;r) = V_{\rm v}(E) \, f(R_{\rm v}, a_{\rm v}; r).
\label{n.13a}\eeq
\noindent 2) Real surface potential:
\beq
V_{\rm d} (E;r) = V_{\rm d}(E) \, 4 a_{\rm d} \,
g(R_{\rm d}, a_{\rm d}; r).
\label{n.13b}\eeq
\noindent 3) Coulomb potential:
\beq
V_{\rm c} (r) = Z e^2 \left\{
\begin{array}{ll}
\displaystyle{ \frac{1}{2 R_{\rm c}} \left( 3 - \frac{r^2}{R_{\rm c}^2}
\right)} \rule{5mm}{0mm}
& \mbox{if $r < R_{\rm c}$,} \\ [5mm]
\displaystyle{\frac{1}{r}} & \mbox{if $r \ge R_{\rm c}$.}
\end{array} \right.
\label{n.13c}\eeq
\noindent 4) Real spin-orbit potential:
\beq
V_{\rm so} (E;r) = V_{\rm so}(E)
\, \frac{1}{r} \, g(R_{\rm so}, a_{\rm so}; r) .
\label{n.13d}\eeq
\noindent 5) Imaginary volume potential:
\beq
W_{\rm v} (E;r) = W_{\rm v}(E) \,
f(R_{\rm w}, a_{\rm w}; r).
\label{n.13e}\eeq
\noindent 6) Imaginary surface potential:
\beq
W_{\rm d} (E;r) = W_{\rm d}(E) \, 4 a_{\rm wd}
\, g(R_{\rm wd}, a_{\rm wd}; r) .
\label{n.13f}\eeq
\\
\noindent 7) Imaginary spin-orbit potential:
\beq
W_{\rm so} (E;r) = W_{\rm so}(E)
\, \frac{1}{r} \, g(R_{\rm wso}, a_{\rm wso}; r).
\label{n.13g}\eeq
The operators ${\bf L}$ and ${\bf S}$ are, respectively, the orbital and
spin angular momenta (both in units of $\hbar$) of the proton. We have
indicated explicitly that the strengths of the potential terms are
functions (usually expressed as polynomials) of the kinetic energy $E$
of the projectile in the L frame.
\end{subequations}

The interaction energy of the proton with the entire atom is obtained by
introducing the screening of the nuclear Coulomb field by the atomic
electrons, \ie,
\beq
V_{\rm at}(r) = \left[ V_{\rm nuc} (r) - V_{\rm c}(r) \right]
+ V_{\rm c} (r) \, \Phi(r).
\label{n.14}\eeq
The terms in square brackets represent the non-Coulombian nuclear
interaction, which has a range of the order of the nuclear radius,
\beq
R_{\rm nuc} \simeq 1.2 \, A^{1/3} \; \mbox{fm.}
\label{n.15}\eeq
The last term is the screened Coulomb interaction that, for $r>R_{\rm
nuc}$, reduces to the atomic form and has a range of the order of the
``atomic radius'',
\beq
R_{\rm at} = Z^{-1/3} a_0,
\label{n.16}\eeq
where $a_0=5.29 \times 10^4$ fm is the Bohr radius.


\subsection{Classical equation of motion and wave equation}

Let us make a digression to discuss the approximations involved in the
calculation of the DCSs in CM. Because nuclear optical-model potentials
are determined by requiring that partial-wave calculations yield results
consistent with experimental data, we are compelled to assume that the
interaction potential $V(r)$ is central. Although this assumption is
valid for non-relativistic projectiles, it becomes inconsistent at high
energies because the Lorentz-FitzGerald contraction destroys the
spherical symmetry of the interaction (see, \eg, Ref.\
\cite{Jackson1975}). We present here a brief derivation of the
relativistic wave equation for the relative motion of the colliding
particles to point out that conventional calculations based on the
central-potential approximation not only neglect the asymmetry of the
potential but also disregard other components of the effective
interaction.

In the CM frame, and assuming a central real potential, the classical
equations of motion of the two particles are
\beq
\dot{\bf p}'_{\rm p} = {\bf F}
\quad \mbox{and} \quad
\dot{\bf p}'_{\rm A} = -{\bf F}
\label{n.17}\eeq
where ${\bf p}'_{\rm p} = {\bf p}'$ and ${\bf p}'_{\rm A}=-{\bf p}'$ are
the linear momenta of the particles, and
\beq
{\bf F} = - \nabla V(r) = - \frac{\d V}{\d r} \, \hat{\bf r},
\label{n.18}\eeq
is the force on the proton (p), which is a function of its position relative
to the atom (A), ${\bf r} = {\bf r}'_{\rm p} - {\bf r}'_{\rm A}$. Dotted
quantities represent time derivatives. The relative velocity, ${\bf v} =
\dot{\bf r}$, is
\beq
{\bf v} = {\bf v}'_{\rm p} - {\bf v}'_{\rm A} =
\frac{c^2 {\bf p}'_{\rm p}}{{\cal W}'_{\rm p}} -
\frac{c^2 {\bf p}'_{\rm A}}{{\cal W}'_{\rm A}} = \left(
\frac{c^2}{{\cal W}'_{\rm p}} + \frac{c^2}{{\cal W}'_{\rm A}} \right)
{\bf p}' ,
\label{n.19}\eeq
where
\beq
{\cal W}'_{\rm p} = \sqrt{m_{\rm p}^2 c^4 + c^2 p'^2} \qquad \mbox{and} \qquad
{\cal W}'_{\rm A} = \sqrt{M_{\rm A}^2 c^4 + c^2 p'^2}
\label{n.20}\eeq
are, respectively, the total energies of the proton and the atom.

By virtue of a {\it vis viva} theorem, the quantity
\beq
s \equiv {\cal W}'_{\rm p} + {\cal W}'_{\rm A} + V(r)
\label{n.21}\eeq
is a constant of motion. Long before the collision,
when the distance $r$ is large and $V(r)=0$, $s$ reduces to the total
energy in CM, ${\cal W}'_{\rm pi}+{\cal W}'_{\rm Ai}$, and we can write
\beq
s^2 = ({\cal W}'_{\rm pi}+{\cal W}'_{\rm Ai})^2 - c^2 ({\bf p}'_{\rm pi} + {\bf
p}'_{\rm Ai})^2,
\label{n.22}\eeq
where the right-hand side is $c^2$ times the square of the total
four-momentum in CM, a Lorentz invariant. Evaluated with the initial
four-momenta in the L frame it reads
\beq
s^2 = \left( m_{\rm p} c^2 + M_{\rm A} c^2 \right)^2 + 2 M_{\rm A}
c^2 E.
\label{n.23}\eeq
As in the non-relativistic study \cite{Goldstein1980}, the value of $s$
determines the function $p'(r)$ as follows. The equality
\beq
s - V(r)
= \sqrt{m_{\rm p}^2 c^4 + p'^2 c^2} + \sqrt{M_{\rm A}^2 c^4 + p'^2 c^2}
\label{n.24}\eeq
implies that
\beq
p'^2(r) = \frac{1}{4 c^2}
\left\{ \left[ s - V(r) \right]^2
+ \frac{(M_{\rm A}^2 - m_{\rm p}^2)^2 c^8}{\left[ s - V(r)
\right]^2} \right\} -
\frac{(M_{\rm A}^2 + m_{\rm p}^2) c^2}{2} \, .
\label{n.25}\eeq
With simple rearrangements, this equation can be cast in the
familiar non-relativistic form
\beq
p'^2(r) = p'^2_{\rm pi} - 2 \mu_{\rm r} V_{\rm ef}(r)
\label{n.26}\eeq
where
\beq
\mu_{\rm r} =  c^{-2}
\frac{{\cal W}'_{{\rm pi}}{\cal W}'_{{\rm Ai}}}
{{\cal W}'_{{\rm pi}}+{\cal W}'_{{\rm Ai}}}\, .
\label{n.27}\eeq
is the relativistic reduced mass and $V_{\rm ef}(r)$ is an effective
potential given by
\beq
V_{\rm ef}(r) = V(r) + V_{\rm r1}(r) + V_{\rm r2}(r),
\label{n.28}\eeq
with
\beq
V_{\rm r1}(r) =
- \frac{V^2(r)}{2 \mu_{\rm r} c^2}
\left( 1 - \frac{3\mu_{\rm r}c^2}{s} \right)
\label{n.29}\eeq
and
\beq
V_{\rm r2}(r) =
\frac{(m_{\rm p}^2 - M_{\rm A}^2)^2 c^6}
{8 \mu_{\rm r} s^2}
\left[ \left( 1 - \frac{V(r)}{s} \right)^{-2} - 1
- 2 \frac{V(r)}{s} - 3 \frac{V^2(r)}{s^2}
\right].
\label{n.30}\eeq
The terms $V_{\rm r1}(r)$ and $V_{\rm r2}(r)$ are corrections to the
interaction potential that account for the effect of relativistic
kinematics. The first term is always attractive and proportional to
$V^2(r)$. Considering the Taylor expansion of the function $(1-x)^{-2}$,
one sees that the second correction term, $V_{\rm r2}(r)$, is of order
$(V/s)^3$, and it vanishes when the projectile and the target particles
have the same mass.

The quantum wave equation governing the motion of the projectile
relative to the target atom can be obtained from the classical equation
\ref{n.26}, by introducing the replacement ${\bf p}' \rightarrow - {\rm
i} \hbar \nabla$, and considering that the resulting operators act on
the wave function $\psi({\bf r})$ (see, \eg, Ref. \cite{Messiah1999}).
We thus find the time-independent wave equation for free states
\beq
\left( - \frac{\hbar^2}{2 \mu_{\rm r}} \, \nabla^2
+ V_{\rm ef}(r) \right) \psi({\bf r})
= \frac{p'^2_0}{2\mu_{\rm r}} \, \psi({\bf r}) \, ,
\label{n.34}\eeq
which has the same form as the non-relativistic Schr\"odinger equation
for scattering of a particle with the relativistic reduced mass
$\mu_{\rm r}$ and linear momentum $p'_0$ by the potential $V_{\rm ef}
(r)$.  Hence, the wave function $\psi({\bf r})$ and the scattering DCS
can be evaluated using the methods and approximations of
non-relativistic quantum theory. This formulation ensures that the
calculated DCS approximates the well-known classical result
\cite{Bohr1948, Goldstein1980} in the limit where classical mechanics is
valid.

To justify the wave equation \req{n.34}, let us assume for a moment that
the interaction is purely electrostatic.  When the mass $M_{\rm A}$ of
the target atom tends to infinity, $\beta_{\rm CM} \simeq 0$ and the CM
frame coincides with the L frame, where the potential is strictly
central. Under these circumstances, ${\bf p}'_0 = {\bf p}$, $s \simeq
\infty$,
\beq
\mu_{\rm r} \simeq c^{-2} {\cal W}'_{pi} = \gamma m_{\rm p},
\quad \mbox{with} \quad
\gamma = \sqrt{1 + \left(\frac{p'_0}{m_{\rm p} c}\right)^2} \, ,
\label{n.35}\eeq
\beq
V_{\rm ef}(r) =  V(r) \left[ 1 - \frac{V(r)}{2 \gamma m_{\rm p} c^2}
\right],
\label{n.36}\eeq
and the wave equation becomes
\beq
\left( - \frac{\hbar^2}{2 \gamma m_{\rm p}} \, \nabla^2
+ V(r) \left[ 1 - \frac{V(r)}{2 \gamma m_{\rm p} c^2}
\right] \right) \psi({\bf r}) =
\frac{p'^2_0}{2\gamma m_{\rm p}} \, \psi({\bf r}),
\label{n.37}\eeq
which, as expected, coincides with the Klein-Gordon (or relativistic
Scr\"{o}dinger) equation for free states of a proton with initial
momentum $p'_0$ in the electrostatic potential $V(r)$ (see, \eg, Ref.
\cite{Schiff1968}). Our derivation shows that the second term
in the effective Klein-Gordon potential \req{n.36} has a purely
kinematic origin.

It is worth noticing that, in the energy range of interest for
transport calculations, the de Broglie wavelengths, $\lambda_{\rm dB} =
h/p'_0$, of protons is much smaller than the atomic radius $R_{\rm at}$.
The values of $\lambda_{\rm dB}$ for protons with kinetic energies of 1
MeV and 100 MeV are, respectively, 4.5 fm and 0.44 fm. As a consequence,
the numerical solution of the radial wave equation that determines the
phase-shifts is very difficult. In addition, the partial-wave series
converge extremely slowly, requiring the calculation of a large number
($ \gtrsim 100,000$) of phase-shifts. Since approximate calculation
methods are available for the case of screened Coulomb potentials
(\ie, corresponding to atoms with a point nucleus), we shall consider
screening and nuclear effects separately by using the most accurate
calculation methods that are practicable for each case.


\subsection{Electronic screening}

In the case of scattering of energetic protons by screened atomic
potentials of the type given by Eq.\ \req{n.8}, the relativistic
correction terms $V_{\rm r1}(r)$ and $V_{\rm r2}(r)$, Eqs.\ \req{n.29}
and \req{n.30}, decrease rapidly when $r$ increases. The corresponding
DCS is strongly peaked at small angles, which correspond to large impact
parameters, at which the correction terms are much smaller than the
potential $V(r)$. Numerical calculations show that the effect of these
correction terms on the DCS is limited to large angles (larger than
about 50 degrees for protons with energies up to 1 GeV), where the DCS is
more than ten orders of magnitude smaller than at forward angles. Hence,
for the purposes of proton transport simulations \cite{Moliere1947,
Salvat2013}, the DCS for the potentials \ref{n.5} can be calculated from
the wave equation \req{n.34} with $V_{\rm eff}(r)=V_{\rm scr}(r)$,
\beq
\left( - \frac{\hbar^2}{2 \mu_{\rm r}} \, \nabla^2
+ V_{\rm scr}(r) \right) \psi({\bf r})
= \frac{p'^2_0}{2\mu_{\rm r}} \, \psi({\bf r}) \, .
\label{n.38}\eeq
It is worth noticing that, within this scheme, the definition \req{n.27}
of $\mu_{\rm r}$ accounts for the correct relativistic kinematics. Here
we are disregarding the effect of the spin of the proton, which is
appreciable only at large scattering angles; it will be accounted for
in the calculation of nuclear scattering (see below).

We recall that the DCS for collisions of protons with a bare point nucleus
is given by the relativistic Rutherford formula,
\beq
\frac{\d \sigma_{\rm R}}{\d \Omega'} = \frac{ \left( 2 \mu_{\rm r} Z e^2
\right)^2}{(\hbar q')^4}
\label{n.39}\eeq
where
\beq
\hbar q' = \left| {\bf p}'_{\rm pi} - {\bf p}'_{\rm pf} \right| = 2 p'_0
\sin(\theta/2)
\label{n.40}\eeq
is the momentum transfer. Because the effect of screening decreases when
the scattering angle increases (\ie, when the classical impact parameter
decreases), the DCS calculated from Eq.\ \req{n.38} tends to the
Rutherford DCS at large angles.

As indicated above, the smallness of the proton wavelength makes the
partial-wave calculation of the DCS for scattering by the screened
Coulomb potential unfeasible. A practical approach adopted in the
previous version of {\sc penh} \cite{Salvat2013} is to use DCSs
calculated with the eikonal approximation \cite{Moliere1947, Schiff1968,
Wallace1971}, in which the phase of the scattered wave is obtained from
a semi-classical approximation to the scattering wave function under the
assumption of small angular deflections of the projectile. To simplify
calculations with the eikonal approximation, the atomic mass $M_{\rm A}$
is set equal to the average atomic mass of the isotopes of the element
weighted by their natural abundances \cite{Wang2012}.

The DCS for scattering by a screened Coulomb potential resulting from
the eikonal approximation is
\beq
\frac{\d \sigma_{\rm scr}}{\d \Omega'} =
\left| f_{\rm eik} (\theta') \right|^2
\label{n.41}\eeq
The function $f_{\rm eik}(\theta')$ is the eikonal scattering
amplitude at the polar scattering angle $\theta'$ for a particle of mass
$\mu_{\rm r}$ and momentum $p'_0 = \hbar k$. It is given by
\beq
f_{\rm eik}(\theta') = - {\rm i} \, k
\int_0^\infty J_0(q'b) \left\{
\exp [{\rm i} \chi(b)]-1\right\} b \, \d b
\label{n.42}\eeq
where $\hbar q$ is the momentum transfer,
$J_0(x)$ is the Bessel function of the first kind and zeroth
order, and $\chi(b)$ is the eikonal phase for projectiles incident with
impact parameter $b$ (including the first-order Wallace correction
\cite{Wallace1971}),
\beq
\chi(b) =
-\, \frac{2 \mu_{\rm r}}{\hbar^2 k} \int_{b}^\infty
V_{\rm scr}(r) \left\{ 1 + \frac{\mu_{\rm r}}{\hbar^2 k^2}
\left[ V_{\rm scr}(r) + r \, \frac{\d
V_{\rm scr}(r)}{\d r} \right] \right\} \frac{ r\, \d r}{\sqrt{r^2 -b^2}}.
\label{n.43}\eeq
For the potential \req{n.8} with a screening
function of the type \req{n.9}, the integral can be solved
analytically \cite{ZeitlerOlsen1967, Salvat2013}, giving
\beq
\chi(b) = - \frac{2\mu_{\rm r} Z e^2 }{\hbar^2 k} \sum_i A_i \left\{
K_0(a_i b) - \frac{\mu_{\rm r} Z e^2}{\hbar^2 k^2} \sum_j A_j a_j
K_0[(a_i+a_j)b] \right\}
\label{n.44}\eeq
where $K_0(x)$ is the modified Bessel function of the second kind and
zeroth order. Thus, the eikonal scattering amplitude can be evaluated by
means of a single quadrature. The numerical calculation is made easier
by a transformation of the integrand due to Zeitler and Olsen
\cite{ZeitlerOlsen1967}, also used in Ref. \cite{Salvat2013}, which
gives
\beq
f_{\rm eik}(\theta') = - \, \frac{k}{q'} \int_0^\infty
J_1(q'b) \frac{\d \chi(b)}{\d b} \exp[{\rm i} \chi(b)] \, b \, \d b,
\label{n.45}\eeq
with
\beqa
\frac{\d \chi(b)}{\d b} &=&
\frac{2 \mu_{\rm r}\, Ze^2}{\hbar^2 k} \,
\sum_i A_i \left\{ a_i K_1(a_i b) \rule{0mm}{8mm}\right.
\nonumber \\ [2mm]
&& \mbox{} \left. \rule{10mm}{0mm}
- \frac{\mu_{\rm r} \, Ze^2}{\hbar^2 k^2} \sum_j A_j a_j
(a_i+a_j)K_1[(a_i+a_j)b] \right\},
\label{n.46}\eeqa
where $J_1(x)$ is the Bessel function of the first kind and first order
and $K_1(x)$ is the modified Bessel function of the second kind and
first order.

The eikonal approximation is expected to be valid for scattering angles
up to about $(k R_{\rm at})^{-1}$ \citep{Moliere1947}. However, numerical
calculations indicate that the approximation yields fairly
accurate DCSs, practically coincident with those obtained from classical
calculations up to much larger angles, of the order of\footnote{The
limiting angle given in Ref.\ \cite{Salvat2013} is not correct.}
\beq
\theta'_{\rm eik} =
\min \left\{ \frac{200}{kR_{\rm a}}, 0.1 \pi \right\}\, .
\label{n.47} \eeq
For still larger angles the calculation loses validity and presents
numerical instabilities. Following \cite{Salvat2013}, the DCS for angles
larger than $\theta_{\rm eik}$ is obtained from the expression
\beq
\frac{\d \sigma_{\rm scr}}{\d \Omega'} =
\left(\frac{2 \mu_{\rm r} Z e^2}{\hbar^2}\right)^2
\frac{1}{\left[ A + B q'^{2/3} + C q'^{4/3} +
q'^2\right]^2}\, .
\label{n.48}\eeq
with the coefficients $A$, $B$ and $C$ obtained by matching the
calculated numerical values of the eikonal DCS and its first and second
derivatives at $\theta'=\theta'_{\rm eik}$. The ratio of the calculated
DCS to the Rutherford DCS,
\beq
F_{\rm scr}(\theta') = \frac{\d \sigma_{\rm scr}}{\d \Omega'}
\left( \frac{\d \sigma_{\rm R}}{\d \Omega'} \right)^{-1},
\label{n.49}\eeq
which measures the effect of screening, approximates unity at large
angles.


\subsection{Nuclear effects in elastic collisions}

The DCS for collisions of protons with bare point nuclei are calculated
from the partial-wave solution of the wave equation
\beq
\left( - \frac{\hbar^2}{2 \mu_{\rm r}} \, \nabla^2
+ V_{\rm nuc}(r) \right) \psi({\bf r})
= \frac{p'^2_0}{2\mu_{\rm r}} \, \psi({\bf r}).
\label{n.50}\eeq
The phenomenological optical-model potentials are set by requiring that
the DCSs obtained in this way agree with available experimental data.

As the  potential \req{n.12} contains spin-orbit terms, the wave
function is a two-component spinor, and scattering observables are
determined by the direct and spin-flip scattering amplitudes, which are
obtained in terms of the phase shifts $\delta_{\ell j}$ of spherical
waves with orbital and total angular momenta $\ell$ and $j$,
respectively. Calculations are performed by using the Fortran
subroutine package {\sc radial} \cite{SalvatFernandezVarea2019}, which
implements a robust power series solution method that effectively avoids
truncation errors and yields highly accurate radial functions. The
reduced radial functions, $P_{\ell j}(r)$ are the regular solutions of
the radial wave equation
\beq
- \frac{\hbar^2}{2 \mu} \, \frac{\d^2}{\d r^2} P_{\ell j}(r)+
\left( V_{\ell j} (r)  + \frac{\hbar^2}{2 \mu} \frac{\ell(\ell+1)}{r^2}
\right) P_{\ell j}(r) = \frac{p'^2_0}{2\mu} P_{\ell j}(r)
\label{n.51}\eeq
with the ``radial'' potential
\beqa
V_{\ell j} (r) &=&
V_{\rm v}(E;r) + V_{\rm d}(E;r) + V_{\rm c}(r)
+ {\rm i} \left[ W_{\rm v} (E;r)+ W_{\rm d} (E;r)\rule{0mm}{3.5mm} \right]
\nonumber \\ [2mm]
&& \mbox{}
+ \left[ V_{\rm so}(E;r) + {\rm i} \, \, W_{\rm so}(E;r)\rule{0mm}{3.5mm}
\right]
 \frac{1}{2} \left( j(j+1) -
\ell (\ell +1) - \frac{3}{4} \right). \rule{10mm}{0mm}
\label{n.52}\eeqa
The radial functions are normalized so that
\beq
P_{\ell j}(r)
\begin{array}[t]{c} \sim \\ [-2mm] \scriptstyle{ r \rightarrow \infty}
\end{array}
\sin \left(kr - \ell \frac{\pi}{2} - \eta \ln (2kr) +  \Delta_\ell +
\delta_{\ell j}
\right),
\label{n.53}\eeq
where
\beq
\eta = \frac{Z e^2 \mu_{\rm r}}{\hbar^2 k}
\label{n.54}\eeq
is the Sommerfeld parameter,
\beq
\Delta_\ell = \arg \left( \ell+1+{\rm i} \eta \right),
\label{n.55}\eeq
is the Coulomb phase shift, and $\delta_{ \ell j}$ is the complex
``inner'' phase shift, which is caused by the finite-range component of
the potential. The inner phase shift is determined by integrating the
radial equation from $r=0$ outwards up to a radius $r_{\rm m}$ larger
than the range of the nuclear interaction, and matching the numerical
solution at $r_{\rm m}$ with a linear combination of the regular and
irregular Coulomb functions. In the following the inner phase shifts are
denoted by the abridged notation $\delta_{\ell a}$ with $a={\rm sign}(j-\ell)$,
\ie, $\delta_{\ell +} \equiv \delta_{\ell,j=\ell+1/2}$ and $\delta_{\ell
-} \equiv \delta_{\ell,j=\ell-1/2}$.

\begin{figure}[h!] \begin{center}
\includegraphics*[width=7.5cm]{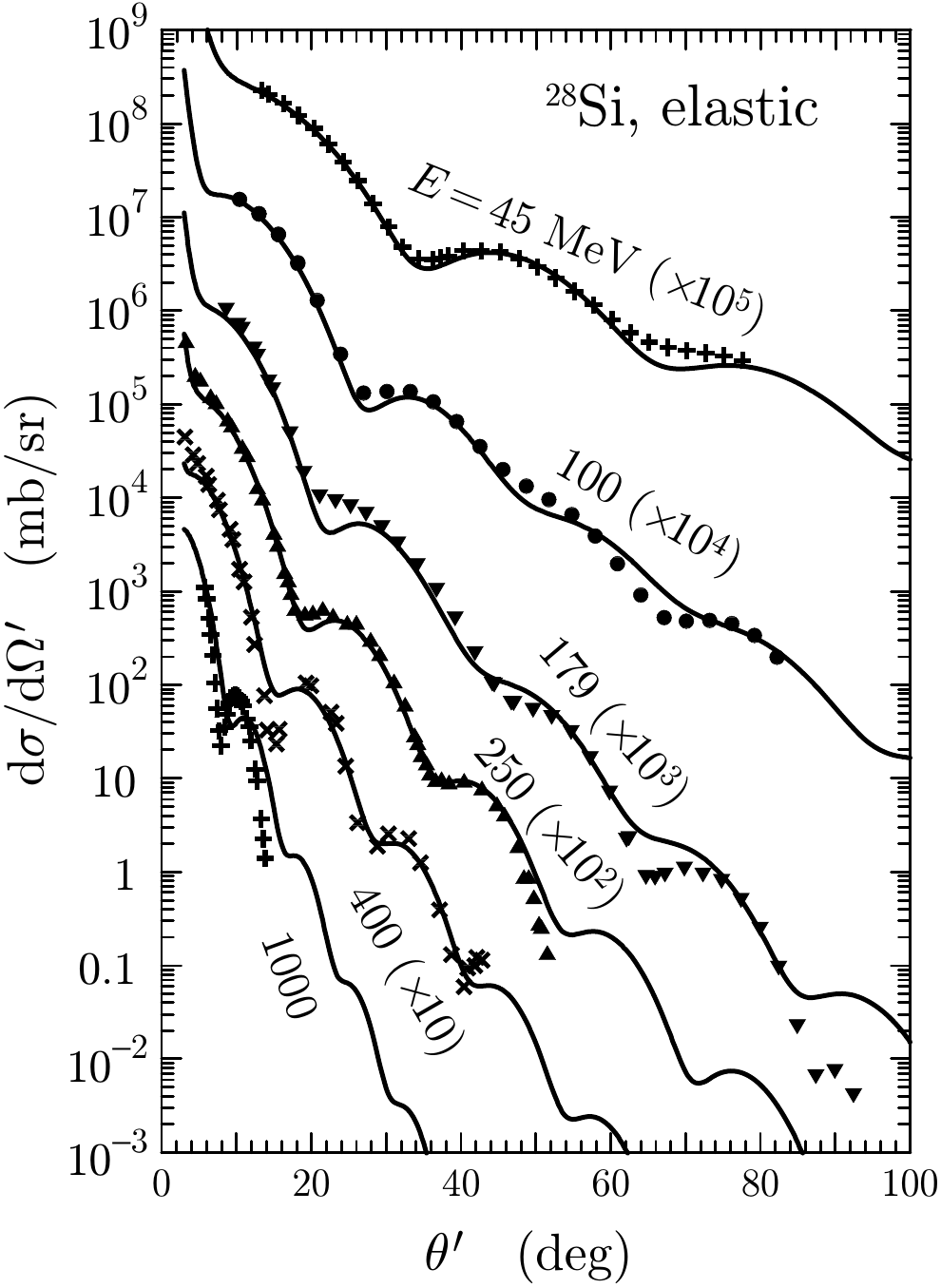}
\caption{Calculated DCSs for protons of the indicated laboratory
kinetic energies (in MeV) with $^{28}$Si nuclei, as functions of the
scattering angle in CM. For the sake of visibility, DCSs are multiplied
by the indicated powers of 10. The symbols represent experimental data
from Refs.\ \cite{Alkhazov1975, Nakamura1983, Olmer1984, Hicks1988},
taken from the EXFOR database (\url{https://www-nds.iaea.org/exfor/}).
\label{fig-28Si}}
\end{center}\end{figure}

For spin-unpolarized projectiles, the elastic DCS per unit solid angle
in the CM frame is given by
\beq
\frac{\d \sigma_{\rm nuc}}{\d \Omega'} =
\left| f(\theta') \right|^2 + \left| g(\theta') \right|^2.
\label{n.56}\eeq
where
\begin{subequations}
\label{n.57}
\beqa
f(\theta') &=& f_{\rm Coul}(\theta') +
\frac{1}{2 {\rm i} k} \sum_{\ell}
{\rm exp}( 2 {\rm i} \Delta_{\ell}) \left[\rule{0mm}{4mm}
(\ell+1) \left( S_{\ell +} - 1
\rule{0mm}{3.5mm}\right) \right.
\nonumber \\ [2mm]
&& \left. \mbox{}
+ \ell \left( S_{\ell -} - 1
\rule{0mm}{3.5mm}\right)\rule{0mm}{4mm}
\right] \, P_\ell(\cos\theta')
\label{n.57a}\eeqa
and
\beq
g(\theta') = \frac{1}{2 {\rm i} k}
\sum_{\ell}  {\rm exp}( 2 {\rm i} \Delta_{\ell})
\left( S_{\ell -} - S_{\ell +}
\rule{0mm}{3.5mm}\right) \, P_\ell^1(\cos\theta'),
\label{n.57b}\eeq
\end{subequations}
are the direct and spin-flip scattering amplitudes, respectively. In these
partial-wave expansions, $P_\ell(\cos\theta')$ and $P_\ell^1(\cos\theta')$
are Legendre polynomials and associated Legendre functions of the first
kind \cite{Olver2010}, respectively,
\beq
S_{\ell a} = \exp(2 {\rm i} \delta_{\ell a}),
\label{n.58}\eeq
are the nuclear parts of the $S$-matrix elements, and
\beq
f_{\rm Coul}(\theta') = - \eta \, \frac{\exp\{ 2{\rm i}\Delta_0
- {\rm i} \eta \ln[\sin^2(\theta'/2)]\}}{2k\sin^2(\theta'/2)}
\label{n.59}\eeq
is the Coulomb scattering amplitude. In the present calculations, the
global optical-model potential of Koning and Delaroche
\cite{KoningDelaroche2003} was adopted. Because the potential parameters
vary smoothly with the atomic number and the mass number, the DCS for
collisions with nuclei of the element of atomic number $Z$ was obtained
as the average of the DCSs for the various stable isotopes of the
element, weighted by the natural abundances of the isotopes
\cite{Wang2012}. Figure \ref{fig-28Si} compares the calculated DCSs in
CM for elastic collisions of protons with $^{28}$Si nuclei and
experimental data, for the indicated kinetic energies of the protons in
L. Although the adopted global model potential is considered to be valid
only for energies up to 200 MeV \cite{KoningDelaroche2003}, it is found
to give a reasonable description of the collisions also for much higher
energies.

Since protons with impact parameters larger than $R_{\rm nuc}$ only feel
the Coulomb force, the calculated DCS reduces to the Rutherford DCS at
small angles. The effect of nuclear interactions is usually
exhibited by giving the ratio of the calculated or measured DCS to the
Rutherford DCS,
\beq
F_{\rm nuc}(\theta') =
\frac{\d \sigma_{\rm nuc}}{\d \Omega'}
\left( \frac{\d \sigma_{\rm R}}{\d \Omega'} \right)^{-1},
\label{n.60}\eeq
which approaches unity at small angles.


\subsection{Atomic DCS and total cross section}

\begin{figure}[h!] \begin{center}
\includegraphics*[width=6.5cm]{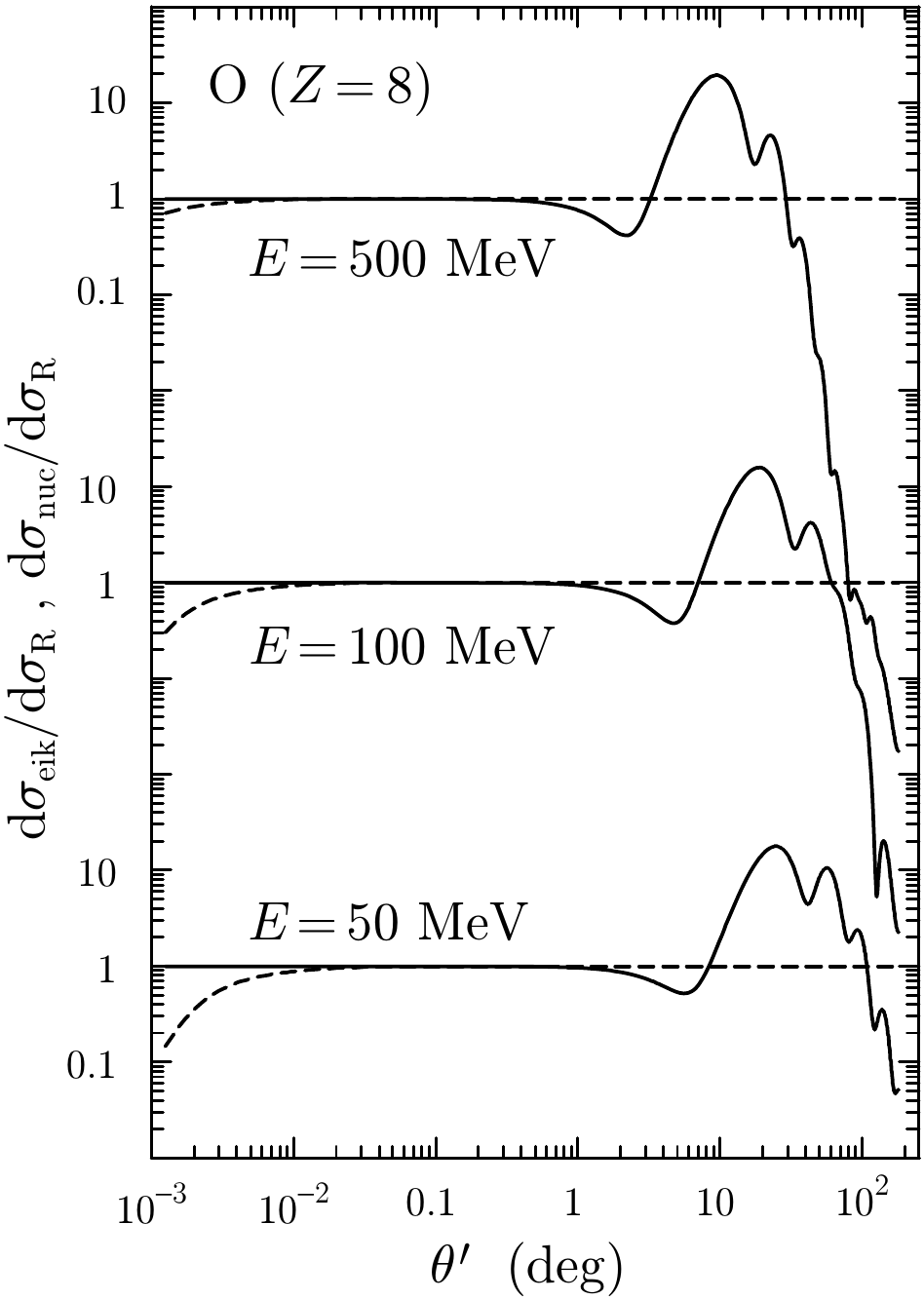} \rule{2mm}{0mm}
\includegraphics*[width=6.65cm]{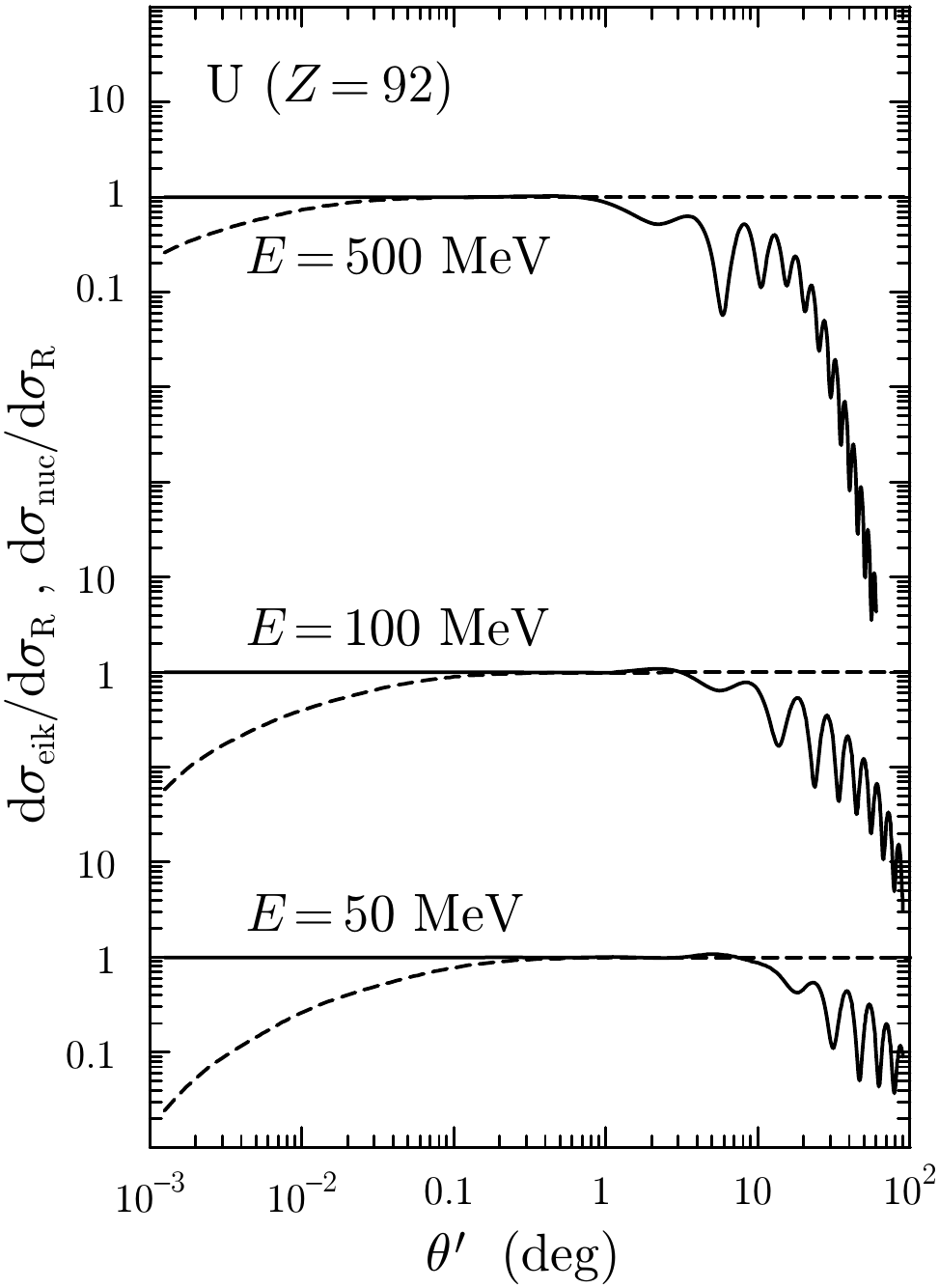}
\caption{Ratios of calculated DCSs for collisions of protons having the
indicated laboratory energies with oxygen and uranium. Dashed curves
represent the $F_{\rm scr}(\theta')$ ratio for the screened atomic
potential, Eq.\ \req{n.49}; solid curves represent the $F_{\rm
nuc}(\theta')$ ratio for the nuclear potential, Eq.\ \req{n.60}.
\label{fig1}}
\end{center}\end{figure}

To account for screening and nuclear effects simultaneously, we consider
the two effects as modifications of Rutherford scattering. It is a
well-known fact
(see, \eg, Ref.\ \cite{Jackson1975}) that screening affects the
DCS only at small angles (impact parameters much larger than $R_{\rm
nuc}$), while nuclear effects are only appreciable at large
scattering
angles (impact parameters of the order of, or less than $R_{\rm nuc}$).
The situation is illustrated in Fig.\ \ref{fig1} which shows ratios of
DCSs in CM calculated with the screened potential \req{n.49} and with
the nuclear potential \req{n.60} to the Rutherford DCS for collisions of
protons having various kinetic energies in L with atoms and nuclei of
oxygen and uranium.

The atomic DCSs adopted in the present simulations were calculated by
assuming that screening and nuclear effects do not interfere because, as
shown Fig.\ \ref{fig2}, they are appreciable in different angular
ranges. That is, we set
\beq
\frac{\d \sigma}{\d \Omega'} =
\frac{\d \sigma_{\rm R}}{\d \Omega'}  \, F_{\rm scr}(\theta') \,
F_{\rm nuc}(\theta').
\label{n.61}\eeq
As these DCSs differ from the ones used in the previous version of
{\sc penh} \cite{Salvat2013} by only the nuclear corrections factor
$F_{\rm nuc}(\theta')$, differences between simulations with the two DCS
sets are appreciable only at angles larger than a few degrees.

Due to the finite size of the nucleus, the DCS has a oscillatory
structure with the number and positions of the minima varying
continuously with the energy of the projectile. The variation of this
structure is weaker when the DCS is considered as a function of the
momentum transfer $q'$. We have generated an extensive numerical
database of DCSs, Eq.\ \req{n.61}, for elastic collisions of unpolarized
protons with atoms of the elements with atomic number from $Z=1$
(hydrogen) to $Z=99$ (einsteinium) and that cover the interval
of
laboratory energies from 100 keV to 1 GeV. To minimize the size of the
database, the DCS is tabulated as a function of the variable $(cq')^2$,
with about 200 grid points unevenly spaced to reduce interpolation
errors, for a logarithmic grid of proton energies with 36 points per
decade.

It is worth mentioning that these calculated DCSs differ from those used
in the 2013 version of {\sc penh} \cite{Salvat2013} not only in the
consideration of nuclear effects in elastic collisions. The old DCSs
were computed from the Klein-Gordon wave equation of a proton with the
CM momentum $p'_0$, \ie, from Eq.\ \req{n.37}. That is, in the old code,
the relativistic reduced mass $\mu_{\rm r}$ in the Klein-Gordon equation
was replaced with the relativistic mass
\beq
\gamma m_{\rm p} = \frac{\sqrt{m_{\rm p}^2 c^4 + c^2 p'^2_0}}{m_{\rm p}
c^2} \, m_{\rm p}.
\label{n.62a}\eeq
In the present version of {\sc penh} the kinematics of relativistic
collisions is described correctly, and we only retain the apparently
unavoidable approximation of assuming a central interaction in the CM
frame.


\subsection{Random sampling}

The simulation of elastic collisions is performed by using the same
strategy as in the {\sc penelope} code and in the previous version of
{\sc penh} \cite{Salvat2015, Salvat2013}. Mean free paths
and other energy-dependent quantities are obtained by log-log linear
interpolation of tables, prepared at the start of the simulation run,
with a logarithmic grid of 200 laboratory energies $E_i$ that covers the
interval of interest. The angular distribution of scattered protons in
CM
\beq
p(E_i, \theta') = \frac{1}{\sigma(E_i)} \, \frac{\d \sigma(E_i, \theta')}{\d
\Omega'}\, ,
\label{n.62}\eeq
is tabulated at the same grid energies.

The scattering angle $\theta'$ of a projectile proton with laboratory
energy $E$ in the interval $(E_i, E_{i+1}]$ is sampled from the
distribution
\begin{subequations}
\label{n.63}
\beq
p(E,\theta') = \pi_i \, p(E_i,\theta') + \pi_{i+1} \, p(E_{i+1},\theta')
\label{n.63a}\eeq
with
\beq
\pi_{i} = \frac{ \ln E_{i+1} - \ln E}{\ln E_{i+1} - \ln E_{i}}
\qquad \mbox{and} \qquad
\pi_{i+1} = \frac{ \ln E - \ln E_i}{\ln E_{i+1} - \ln E_{i}}
\label{n.63b}\eeq
\end{subequations}
which is obtained from the tabulated distributions by linear
interpolation in $\ln E$. The sampling is performed by using the
composition method: \\
1) select the value of the index $k=i$ or $i+1$, with respective point
probabilities $\pi_i$ and $\pi_{i+1}$, and \\
2) sample $\theta'$ from the distribution $p(E_k,\theta')$. \\
With this interpolation by weight method, $\theta$ is generated by
sampling from only the distributions at the grid energies $E_i$.  This
sampling is performed by the inverse transform method by using the RITA
(rational interpolation with aliasing) algorithm \cite{GarciaTorano2019,
Salvat2015}. The required sampling tables are prepared by the program at
the start of the simulation run.

The energy and polar scattering angle of the projectile in the
L frame are obtained by a Lorentz boost, and recalling that ${\cal
W}'_{\rm pf} = {\cal W}'_{\rm pi}$,
\beq
E_{\rm pf} = \gamma_{\rm CM} \left( W'_{\rm pf} +
\beta_{\rm CM}  \, cp'_0 \, \cos\theta' \right) - m_{\rm p} c^2
\label{n.64}\eeq
and
\beq
\cos\theta_{\rm p} = \frac{\tau_{\rm p} + \cos\theta'}{
\sqrt{(\tau_{\rm p} + \cos\theta')^2 + \gamma_{\rm CM}^{-2} \sin^2
\theta'}},
\label{n.65}\eeq
where
\beq
\gamma_{\rm CM} = \frac{1}{\sqrt{1 - \beta_{\rm CM}^2}}
\label{n.66}\eeq
and $\tau_{\rm p}$ is the ratio of speeds of the CM and of the scattered
proton, $v'_{\rm P} = c^2 p'_0/{\cal W}'_{\rm pf}$,
\beq
\tau_{\rm p} = \frac{v_{\rm CM}}{v'_{\rm p}} = \sqrt{ \left(
\frac{m_{\rm p}}{M_{\rm A}} \right)^2 (1-\beta_{\rm CM})^2 +
\beta_{\rm CM}}.
\label{n.67}\eeq
The recoil energy of the target atom, $E_{\rm A}=E-E_{\rm p}$, is
assumed to be deposited locally, except when the target is a hydrogen
atom, in which case we consider the recoiling target nucleus as a
secondary proton with initial energy $E_{\rm A}$ and direction in the
scattering plane with polar angle given by
\beq
\cos\theta_{\rm A} = \frac{1-\cos\theta'}{\sqrt{(1-\cos\theta')^2
+\gamma_{\rm CM}^{-2} \sin^2 \theta'}}.
\label{n.68}\eeq


\section{Inelastic collisions}

The slowing down of fast protons is caused primarily by electronic
excitations of the material, usually referred to as inelastic collisions.
The program {\sc penh} simulates electronic excitations by using DCSs
derived by means of the relativistic first-order Born approximation
\cite{Salvat2013}, which are expressed as the product of a kinematic
factor and the generalized oscillator strength (GOS), a function of the
energy transfer $W$ and the momentum transfer $q$. The GOS is
approximated by the Sternheimer-Liljequist (SL) model, which describes
excitations of each atomic subshell by a single $\delta$-oscillator,
with strength equal to the number of electrons in the subshell and
resonance energy determined to reproduce the adopted empirical value of
the mean excitation energy $I$ of the material. This GOS model
is essentially the same as the model used in {\sc penelope} for
projectile electrons and positrons \cite{Salvat2015}. It gives fairly
realistic stopping powers for a wide range of proton energies and
it partially accounts for the correlation between energy and momentum
transfers with the angular deflection of the projectile.

The subshell ionization cross sections resulting from the SL GOS are not
sufficiently accurate for describing the production of x rays by proton
impact, which is important for, \eg, the simulation of
proton-induced
x-ray spectra. This deficiency is remedied by considering more realistic
ionization cross sections for inner subshells with binding energies
higher than 50 eV. These cross sections were obtained numerically using
GOSs calculated by considering an independent-electron model with the
Dirac-Hartree-Fock-Slater potential of free atoms \cite{BoteSalvat2008,
Salvat2013}. The DCS resulting from the SL model are renormalized to
reproduce the numerical ionization cross sections. This renormalization
affects only the proton mean free path for excitations of inner
subshells, it does not alter the distribution of energy and momentum
transfers of the SL model. At the same time, the mean free paths for
excitations of outer sub-shells are renormalized to keep the total
stopping power unaltered. This scheme provides a consistent description
of the production of secondary electrons and x rays in inelastic
collisions, with proper account of the correlations of the emission and
the energy loss of the projectile, without complicating the sampling
algorithm.

Because of the simplicity of the GOS, the SL model is not valid for
protons with very low energies. A great deal of experimental information
on the stopping of protons in materials has been collected over the
years and is available from the IAEA web site
\cite{Montanari2017}.
These empirical data have been combined with theoretical models to set
up reference tables of proton stopping powers of various materials
\cite{ICRU49}, which can be produced and downloaded by running the
program {\sc pstar} \cite{Berger1992} from the NIST web site. The code
{\sc penh} permits the stopping power of protons in a material to be
defined
by means of a table of values for a given grid of energies. When such a
table is provided, the DCSs of outer subshells, obtained as summarized
above, are further renormalized to reproduce the input stopping power.
Thus, {\sc penh} is capable of simulating inelastic collisions by using
the most reliable stopping powers available, together with a realistic
description of the energy-loss and the angular deflection of the
projectile proton.

\allowdisplaybreaks{

\section{Nuclear reactions\label{sec3}}

A fast proton may undergo an inelastic interaction with a nucleus of the
various isotopes present in the material. As a result of this
interaction, a number of reaction products (neutrons, protons, tritons,
alphas, prompt gammas, \ldots, and residual nuclei) may be emitted.

As indicated in the Introduction, we simulate proton-induced nuclear
reactions by using information from nuclear databases in the standard
ENDF-6 format \citep{ENDF2018}, which allows the use of reaction data
from either the US ENDF/B evaluated data libraries
(\url{https://t2.lanl.gov/nis/data.shtml}) or other ENDF-formatted data
libraries such as the TENDL-2019
(\url{https://tendl.web.psi.ch/tendl_2019/tendl2019.html}) \citep{Koning2019}. The ICRU
Report 63 \cite{ICRU63} provides an extension to 250 MeV of the
ENDF/B-VI Release 6 library for the major
isotopes of H, C, N, O, Al, Si, P, Ca, Fe, Cu, W, and Pb (the original
ENDF/B libraries extended only up to 150 MeV).
For a given isotope $^AZ$, these
data files provide the reaction cross section $\sigma_{\rm nr}(E)$ and a
statistical description of the products emitted in a reaction, as
functions of the kinetic energy $E$ of the proton in the laboratory
reference frame. For each reaction the ENDF-6 formatted files give a list of
released product types, the average number of products of each type, and
the corresponding angular and energy distributions. This description
ignores correlations between various products as well as specific
reaction channels and, consequently, energy and baryon number are
conserved on average. It represents the results that would be obtained from
measurements with a detector that counts products of one given type that
are emitted with energies and directions within the acceptance limits of
the device.  Although partial, this information is sufficient for
describing the influence of nuclear reactions on the dose distributions
from proton beams. Because the tables cover only proton energies $E$ up
to a maximum value $E_{\rm max}$ of the order of 250 MeV, the program
may have to extrapolate to higher energies.

The reaction cross section determines the probability of occurrence of
nuclear reactions per unit path length of the transported protons. Its
inverse is the reaction mean free path
\beq
\lambda_{\rm nr}(E) = {\cal N}_{ZA} \sigma_{\rm nr}(E)
\label{n.69}\eeq
where ${\cal N}_{ZA}$ is the number of nuclei per unit volume of the
isotope $^AZ$, and $\sigma_{\rm nr}(E)$ is the corresponding reaction
cross section. For energies higher than $E_{\rm max}$ the reaction cross
section is assumed to be constant, \ie, equal to $\sigma_{\rm nr}(E_{\rm
max})$.

The distribution of reaction products in type, number, energy, and angle
is simulated using the information in the nuclear data files. In the
rest of the present Section we use the notation of the ENDF-6 Formats
Manual \cite{ENDF2018} with all kinematic quantities pertaining to the
emission channel referred to the CM frame. We assume that the projectile
${\rm a}$ moves initially in the direction of the $z$ axis with kinetic
energy $E_{\rm a}$. Let us consider a reaction
\beq
{\rm a}+{\rm A} \rightarrow {\rm C} \rightarrow {\rm b} + {\rm B}
\label{n.70}\eeq
in which the product ${\rm b}$ is emitted with kinetic energy $E'_{\rm
b}$ in a direction with polar angle $\theta'_{\rm b}$, leaving a residual
nucleus ${\rm B}$. Azimuthal symmetry of emission is assumed, and the
direction of the product is characterized by the polar direction cosine
$\cos \theta'_{\rm b}$. The emission of product ${\rm b}$ is
characterized by the double-differential production cross section,
\beq
\frac{\d^2 \sigma_{\rm b} (E_{\rm a})}{\d E'_{\rm b} \; \d \Omega'_{\rm b}}
= \frac{1}{2\pi} \, \sigma_{\rm nr}(E_a) \, y_{\rm b} \,
f_{\rm b}(E_{\rm a}; E'_{\rm b}, \cos \theta'_{\rm b})
\label{n.71}\eeq
where $y_{\rm b}$ is the product yield or multiplicity, and $f_{\rm b}$ is the
normalized energy-direction distribution,
\beq
\int \d E'_{\rm b} \int \d \cos \theta'_{\rm b} \, f_{\rm b}(E_{\rm a};
E'_{\rm b},\cos\theta'_{\rm b}) = 1.
\label{n.72}\eeq
The energy spectrum of the reaction product ${\rm b}$,
\beq
f_{\rm b}(E_{\rm a}; E'_{\rm b}) \equiv \int \d \cos \theta'_{\rm b} \,
f_{\rm b}(E_{\rm a};
E'_{\rm b}, \cos \theta'_{\rm b})
\label{n.73}\eeq
is given as a histogram defined by means of a partition $0 \le E'_1 < E'_2
< \cdots < E'_{N+1}=E'_{\rm b,max}$ of the interval of product
energies and the value $f_{\rm b}(E_{\rm a};E'_j)$ of the spectrum in
each bin $(E'_j, E'_{j+1})$. Explicitly, the energy
spectrum is
\beq
f_{\rm b}(E_{\rm a};E'_{\rm b}) = f_{\rm b}(E_{\rm a};E'_{j})
\quad \mbox{if} \; \; E'_{\rm b} \in  (E'_j, E'_{j+1}).
\label{n.74}\eeq
The angular distribution of products ${\rm b}$ emitted with kinetic
energy $E'_{\rm b}$ is
\beq
p(E_{\rm a}, E'_{\rm b}; \cos \theta'_{\rm b}) =
\frac{f_{\rm b}(E_{\rm a}; E'_{\rm b},\cos \theta'_{\rm b})}{f_{\rm b} (E_{\rm a};
E'_{\rm b})}.
\label{n.75}\eeq
Notice that both $f_{\rm b}(E_{\rm a}; E'_{\rm b})$ and $p(E_{\rm a},
E'_{\rm b};\cos \theta'_{\rm b})$ are normalized to
unity.

Prompt gammas are emitted isotropically, that is,
\beq
p(E_{\rm a}, E'_\gamma; \cos\theta'_\gamma) = \frac{1}{2},
\label{n.76}\eeq
where $\theta'_\gamma$ is the polar angle of the gamma-ray direction in
the CM frame.  The ENDF/B and TENDL files specify the distributions of
energy $E_{\rm b}$ and direction $\theta_{\rm b}$ of heavy recoils {\it
in the L frame}. Since heavy recoils are assumed to be locally absorbed
by the simulation program, their angular distribution is irrelevant.

The angular distributions of the light products listed in Table 1 are
represented as Kalbach distributions \cite{Kalbach1988},
\beqa
p_{\rm K} (E_{\rm a}, E'_{\rm b};\cos \theta'_{\rm b}) &=&
\frac{a(E_{\rm a},E'_{\rm b})}{2 \sinh[a(E_{\rm a},E'_{\rm b})]}
\left\{ \cosh[a(E_{\rm a},E'_{\rm b}) \cos\theta'_{\rm b}]
\rule{0mm}{4mm}\right.
\nonumber \\ [2mm]
&& \left. \mbox{}
+ r(E_{\rm a},E'_{\rm b})
\, \rule{0mm}{4mm} \sinh[a(E_{\rm a},E'_{\rm b}) \cos\theta'_{\rm b}]
\right\}
\label{n.77}\eeqa
where $r(E_{\rm a},E'_{\rm b})$ and $a(E_{\rm a},E'_{\rm b})$ are the
pre-compound fraction and the slope parameter, respectively. These
parameters depend on the energy of the projectile, and on the type and
energy of the product; typically $a$ is of the order of unity and $r$
ranges from 0.0 to 1.0. In the ENDF-6 files $r$ is tabulated for the
same set of projectile and product energies, $E_{\rm a}$ and $E'_{\rm
b}$, as the energy spectrum of the product, and we assume that it is
constant within each energy bin of the product spectrum.

The slope parameter $a(E_{\rm a},E'_{\rm b})$ of the Kalbach angular
distribution is calculated on the fly for the given energies of the
projectile and the product. The products are assumed to be emitted
independently of each other through reactions of the type \req{n.70}.
The calculation of $a(E_{\rm a},E'_{\rm b})$ involves the separation
energy of a light particle ${\rm a}$ in a reaction ${\rm a}+{\rm A}
\rightarrow {\rm C}$ or ${\rm C} \rightarrow {\rm a} + {\rm A}$, which
is given by (the numerical coefficients are energies in MeV)
\beqa
S_{\rm aA}^{\rm C} &=& 15.68 \left[ A_{\rm C}- A_{\rm A} \right]
- 28.07 \left( \frac{(N_{\rm C} - Z_{\rm C})^2}{A_{\rm C}}
- \frac{(N_{\rm A} - Z_{\rm A})^2}{A_{\rm A}} \right)
\nonumber \\ [2mm]
&& \mbox{}
- 18.56 \left( A_{\rm C}^{2/3}- A_{\rm A}^{2/3} \right)
+ 33.22 \left( \frac{(N_{\rm C} - Z_{\rm C})^2}{A_{\rm C}^{4/3}}
- \frac{(N_{\rm A} - Z_{\rm A})^2}{A_{\rm A}^{4/3}} \right)
\nonumber \\ [2mm]
&& \mbox{}
- 0.717 \left( \frac{Z_{\rm C}^2}{A_{\rm C}^{1/3}}-
\frac{Z_{\rm A}^2}{A_{\rm A}^{1/3}} \right)
+ 1.211 \left( \frac{Z_{\rm C}^2}{A_{\rm C}}-
\frac{Z_{\rm A}^2}{A_{\rm A}} \right) - I({\rm a}),
\label{n.78}\eeqa
where $Z_J$, $A_J$ and $N_J$ are the charge, mass, and neutron numbers
of the particle $J$, respectively, and $I({\rm a})$ is the energy
required to separate the light particle $a$ into its constituent
nucleons, see Table \ref{tab1}. The slope parameter is given by
\beq
a(E_a,E'_b) = \frac{0.04}{\rm MeV} X_1 +  \frac{1.8\times 10^{-6}}{{\rm
MeV}^3} X_1^3 + \frac{6.7 \times 10^{-7}}{{\rm MeV}^4}
m'_{\rm b} X_3^4
\label{n.79}\eeq
with
\begin{subequations}
\label{n.80}
\beq
X_1 = \min\{ e_{\rm a}, 130 \; {\rm MeV} \} e_{\rm b}/e_{\rm a}
\label{n.80a}\eeq
and
\beq
X_3 = \min\{ e_{\rm a}, 41 \; {\rm MeV} \} e_{\rm b}/e_{\rm a}
\label{n.80b}\eeq
where
\beq
e_{\rm a} = E'_{\rm a} + E'_{\rm A} + S_{\rm aA}^{\rm C},
\qquad
e_{\rm b} = E'_{\rm b} + E'_{\rm B} + S_{\rm bB}^{\rm C}.
\label{n.80c}\eeq
\end{subequations}
The quantities $E'_A$ and $E'_B$ are the kinetic energies of the nuclei in
the reaction \req{n.70}. Since in the CM frame the magnitudes of the linear
momenta of the light particle ${\rm a}$ (${\rm b}$) and the companion
nucleus ${\rm A}$ (${\rm B}$) are equal, we have
\beq
E'_{\rm A} = \sqrt{E'_{\rm a} (E'_{\rm a} + 2 M_{\rm a} c^2)
+ M_{\rm A}^2 c^4} - M_{\rm A}c^2
\label{n.81}\eeq
and
\beq
E'_{\rm B} = \sqrt{E'_{\rm b} (E'_{\rm b} + 2 M_{\rm b} c^2)
+ M_{\rm B}^2 c^4} - M_{\rm B}c^2,
\label{n.82}\eeq
where $E'_{\rm a}$ is the kinetic energy of the projectile proton in the CM
frame,
\beq
E'_{\rm a} = \gamma_{\rm CM} \left(E + m_{\rm p} c^2 - \beta_{\rm CM} c
p \right) - m_{\rm p} c^2.
\eeq
Values of the coefficient $m'_{\rm b}$ for the considered light products
(with mass number up to 4) are listed in Table \ref{tab1}.

\begin{table}[htb!]
\caption{\rm
Numerical values of the coefficients in the definition \req{n.79} of the
slope parameter $a(E_{\rm a},E'_{\rm b})$ of the Kalbach distribution
for a reaction ${\rm a} + {\rm A}
\rightarrow {\rm C} \rightarrow {\rm b} + {\rm B}$.
\label{tab1}}
\vskip 3mm
\begin{center}
\begin{tabular}{|l|c|c|}
\hline \hline
Particle & \multicolumn{2}{|c|}{Coefficients\rule[-2.0mm]{0mm}{6.5mm}}
\\ \hline \hline
Neutron, $n$ & $I(n) = 0$ &
$m'_n = 1/2$\rule[-2.0mm]{0mm}{6.5mm} \\ \hline
Proton, $p$ & $I(p) = 0$ &
$m'_p = 1$\rule[-2.0mm]{0mm}{6.5mm} \\ \hline
Deuteron, $d$ & $I(d) = 2.22 \; {\rm MeV}$ &
$m'_d = 1$\rule[-2.0mm]{0mm}{6.5mm}\\ \hline
Triton, $t$ & $I(t) = 8.48 \; {\rm MeV}$ &
$m'_t = 1$\rule[-2.0mm]{0mm}{6.5mm} \\ \hline
Helion, $^3$He & $I(^3{\rm He}) = 7.72 \; {\rm MeV}$ &
$m'_{\rm 3He} = 1$\rule[-2.0mm]{0mm}{6.5mm} \\ \hline
Alpha, $\alpha$ & $I(\alpha) = 28.3 \; {\rm MeV}$ &
$m'_\alpha = 2$\rule[-2.0mm]{0mm}{6.5mm}
\\ \hline \hline
\end{tabular} \end{center} \end{table}

} 

\subsection{Random sampling of nuclear reactions}

The simulation of a proton-induced nuclear reaction with a given isotope
$^AZ$ proceeds as follows. For each incident energy, $E_{\rm a}$ in L,
the various types ${\rm b}$ of emitted products are identified. The
number $n_{\rm b}$ of products of type ${\rm b}$ is selected randomly,
averaging to the corresponding yield $y_{\rm b}$. For each emitted
product, the initial kinetic energy $E'_{\rm b}$ (in CM) is sampled from the
corresponding energy spectrum $f_{\rm b}(E_{\rm a}; E'_{\rm b})$. The
direction of emission of the product is obtained by sampling the polar
direction cosine $\cos \theta'_{\rm b}$ from the Kalbach distribution
with the corresponding parameters $a(E_{\rm a},E'_{\rm b})$ and $r(E_{\rm
a},E'_{\rm b})$; the azimuthal angle $\phi'_{\rm b}$ is sampled uniformly.
Finally, a Lorentz boost with velocity $v_{\rm CM}$ gives the energies
and directions of the products in the L frame.

We use the grid of projectile proton energies $E_a$ that is read from
the database files. Reaction cross sections are calculated by linear
log-log interpolation. The production cross sections at other energies
are obtained by the interpolation-by-weight method [see Eqs.\
\req{n.63}] and, consequently, the program does random samplings only
for the energies $E_a$ of that grid. This procedure allows the sampling
of discrete random variables by using Walker's aliasing method
\cite{Walker1977}.

\vspace*{2mm}

\noindent {\it a) Number of reaction products.}

\noindent The nuclear data files provide only the yield $y_{\rm b}$,
\ie, the average number of products of type b emitted in a reaction, which
usually is not an integer. In the simulations, the number $n_{\rm b}$ of
b products in each reaction is generated by using the following
sampling formula
\beq
n_{\rm b} = \left\{
\begin{array}{ll}
[y_{\rm b}]+1 & \mbox{if $\xi < y_{\rm b}-[y_{\rm b}]$,}
\\ [1mm]
y_{\rm b} & \mbox{otherwise,}
\end{array} \right.
\label{n.83}\eeq
where $\xi$ is a random number uniformly distributed in the interval
$(0,1)$, and $[y_{\rm b}]$ is the integer part of $y_{\rm b}$.
The distribution of sampled $n_{\rm b}$ values has the correct mean
$y_{\rm b}$ and minimal variance.

\vspace*{2mm}

\noindent {\it b) Energy spectra.}

\noindent The energy $E'_{\rm b}$ of a product is sampled from the histogram
spectrum, Eq.\ \req{n.74}. Random values of $E'_{\rm b}$ are generated by first
selecting the energy bin, $(E'_j, E'_{j+1})$, with point probability
\beq
p_j = (E'_{j+1}-E'_j) f_{\rm b}(E_{\rm a}; E'_j).
\label{n.84}\eeq
This selection is made by means of Walker's \cite{Walker1977} aliasing
method. Although this method requires the preliminary calculation of
cut-off
values and aliases (see, \eg, \cite{Salvat2015}), it is optimal, in the
sense that the generation speed (number of random values generated per
unit CPU time) is independent of the number of bins in the histogram (at
a cost of increasing the required memory storage by about 50 \%).
The
precise value of the product energy $E'_{\rm b}$ is sampled uniformly
within the bin.

\vspace*{2mm}

\noindent {\it c) Kalbach distribution.}

\noindent The angular distribution of the product is given by the
Kalbach distribution \req{n.77}, which we can write in terms of the
parameters $a=a(E_{\rm a},E'_{\rm b})$ and $r=r(E_{\rm a},E'_{\rm b})$ and
the exponential function,
\beq
p_{\rm K}(a,r; x) = \frac{a}{{\rm e}^a - {\rm e}^{-a}}
\left( \frac{{\rm e}^{ax} + {\rm e}^{-ax}}{2}
+ r \frac{{\rm e}^{ax} - {\rm e}^{-ax}}{2} \right),
\label{n.85}\eeq
where $x=\cos\theta'_{\rm b}$. It is convenient to consider the
equivalent expression
\beq
p_{\rm K}(a,r; x) = \frac{1+r}{2}
\left( \frac{a}{{\rm e}^a - {\rm e}^{-a}} \, {\rm e}^{ax} \right)
+ \frac{1-r}{2}
\left( \frac{a}{{\rm e}^a - {\rm e}^{-a}} \, {\rm e}^{-ax} \right),
\label{n.85a}\eeq
which represents the Kalbach distribution as the mixture of two
normalized exponential distributions in $(-1,1)$ with positive weights.
Random values $x$ from the exponential distribution
\beq
p_{\rm exp}(x) = \frac{a}{{\rm e}^a - {\rm e}^{-a}} \, {\rm e}^{ax}
\qquad x \in (-1,1)
\nonumber \eeq
can be generated by using the inverse transform method, which leads to
the sampling formula
\beq
x = \frac{1}{a} \ln \left[ 1 + \xi \left( {\rm e}^{2a}-1 \right) \right]
- 1,
\nonumber \eeq
where $\xi$ is a random number uniform in $(0,1)$. The representation
\req{n.85a} suggests using the following composition algorithm (see, \eg,
\cite{DunnShultis2012}) for generating random values of $x$ from the
Kalbach distribution: \begin{subequations}\\
1) Generate two random numbers $\xi_1$ and $\xi_2$. \\
2) If $\xi_1 < (1+r)/2$, set
\beq
x = \frac{1}{a} \ln \left[ 1 + \xi_2 \left( {\rm e}^{2a}-1 \right) \right]
- 1.
\eeq
3) Else, deliver
\beq
x = - \frac{1}{a} \ln \left[ 1 + \xi_2 \left( {\rm e}^{-2a}-1 \right) \right]
- 1.
\eeq
Notice that this sampling algorithm is exact.
\end{subequations}

Once the energy $E'_{\rm b}$ and the polar direction cosine $\cos\theta'_b$ are
determined in CM, the energy and direction of the product in the L frame
are obtained by applying a Lorentz boost with velocity $-{\bf v}_{\rm
CM}$,
\beq
E_{\rm b} = \gamma_{\rm CM} \left( E'_b + M_{\rm b} c^2 +
\beta_{\rm CM}  \, cp'_{\rm b} \, \cos\theta'_{\rm b} \right) - M_{\rm b} c^2
\label{n.90}\eeq
and
\beq
\cos\theta_{\rm b} = \frac{\tau_{\rm b} + \cos\theta'_{\rm b}}{
\sqrt{(\tau_{\rm b} + \cos\theta'_{\rm b})^2 + \gamma_{\rm CM}^{-2} \sin^2
\theta'_{\rm b}}},
\label{n.91}\eeq
with
\beq
\tau_{\rm b} = \frac{v_{\rm CM}}{v'_{\rm p}}
\label{n.92}\eeq
where
\beq
p'_{\rm b} = c^{-1} \sqrt{E'_{\rm b}(E'_{\rm b} + 2 M_{\rm b} c^2)}
\label{n.93}\eeq
and
\beq
v'_{\rm b} = \frac{c^2p'_{\rm b}}{E'_{\rm b} + M_{\rm b} c^2}
\label{n.94}\eeq
are, respectively, the linear momentum and the velocity of the product
in CM.
The energy and direction of emission of prompt gammas ($M_\gamma=0$,
$v'_\gamma =c$, $p'_\gamma = E'_\gamma/c$, $\tau_\gamma = \beta_{\rm
CM}$), in the L frame are given by
\beq
E_{\gamma} = \gamma_{\rm CM} \left( 1 + \beta_{\rm CM} \,
\cos\theta'_\gamma \right) E'_\gamma
\label{n.90g}\eeq
and
\beq
\cos\theta_\gamma = \frac{\beta_{\rm CM} + \cos\theta'_\gamma}{
\sqrt{(\beta_{\rm CM} + \cos\theta'_\gamma)^2 + \gamma_{\rm CM}^{-2}
\sin^2 \theta'_\gamma}}.
\label{n.91g}\eeq

\begin{figure}[p!] \begin{center}
\includegraphics*[width=9.0cm]{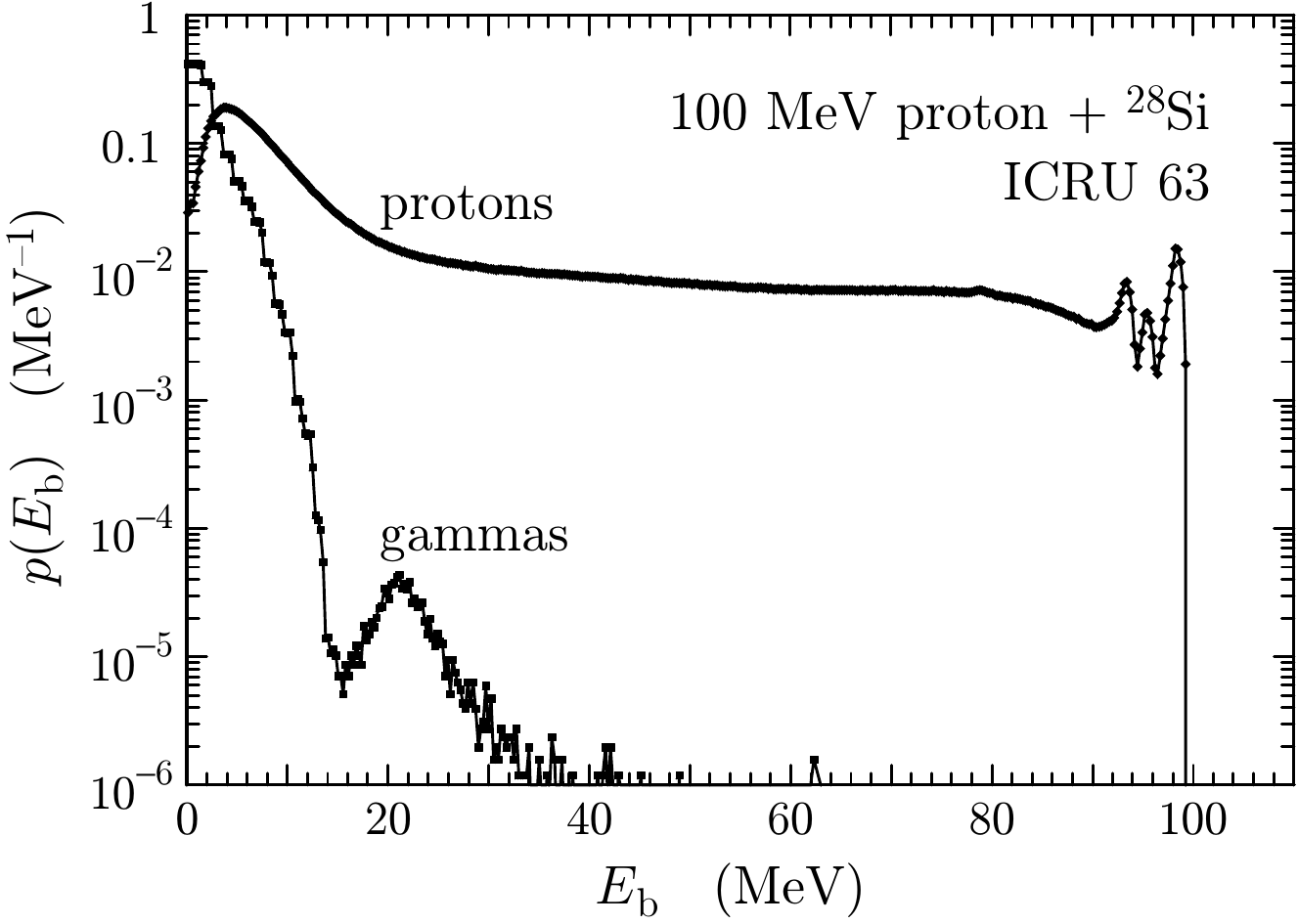} \\ [5mm]
\includegraphics*[width=9.0cm]{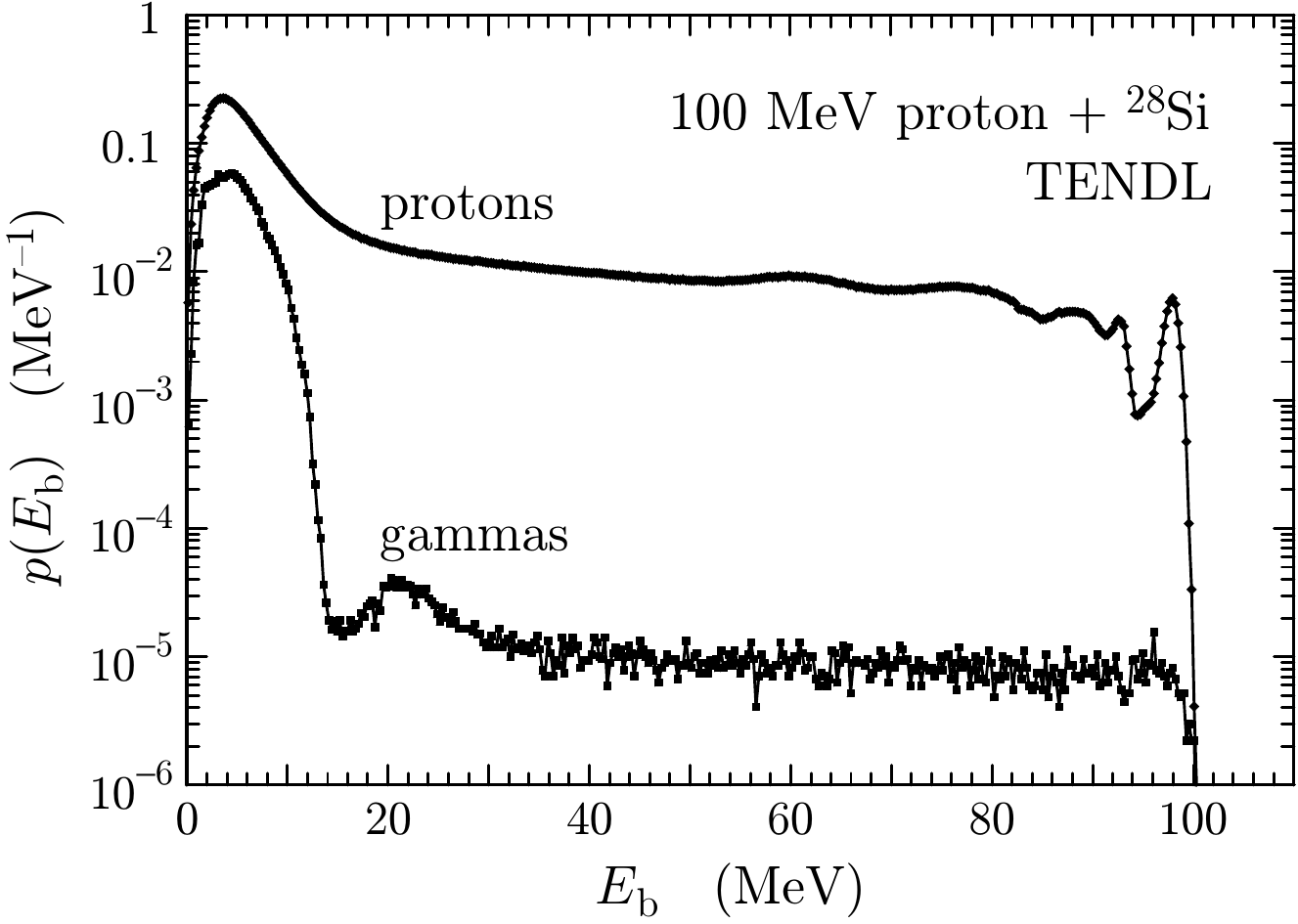}
\caption{Laboratory energy distributions of protons and gammas released in
reactions of 100 MeV protons with $^{28}$Si nuclei, resulting from 10
million random reactions described by using the evaluated ICRU 63 file
(top) and the TENDL file (bottom).
\label{figICRU}}
\end{center}\end{figure}

A practical limitation of basing the simulation of proton-induced
reactions on ENDF-6 formatted files is that these may be available for
only a limited number of nuclides and up to a certain proton energy.
Thus the ICRU 63 Report provides data for only the most abundant
isotopes of 12 elements for energies up to 250 MeV, while ENDF/B
files are
available for about 50 nuclides and $E \le 150$ MeV. On the other hand,
the calculated TENDL files
(\url{https://tendl.web.psi.ch/tendl_2019/tendl2019.html}) cover about
2800 isotopes that live longer than 1 second with proton energies up to
200 MeV. While ENDF/B and ICRU 63 libraries contain evaluated data (\ie,
generated by models with parameters ``locally'' optimized by fitting to
the
available experimental data for each isotope), the TENDL data are
calculated with the TALYS nuclear code system with ``global''
parameters. Evaluated data are expected to be more reliable than
calculated data. Reaction data are generally affected by relatively
large uncertainties and, consequently, data in different libraries may
differ. To illustrate the type of differences that we may find, Fig.\
\ref{figICRU} shows laboratory energy distributions of protons and
gammas emitted in reactions induced by the impact of 100 MeV protons on
$^{28}$Si nuclei, obtained from 10 million sampled reactions, by using
the ICRU 63 and the TENDL data files. The simulated distributions are
normalized so that their integrals equal the average numbers of protons
or gammas released per reaction. In transport simulations we use the
most reliable of the sources at hand, that is, the ENDF-6 files for the
isotopes included in the ICRU 63 Report, the available ENDF/B files for
other isotopes, or TENDL files as a last recourse.

As mentioned above, the reaction data in ENDF-6 files extend up to a
certain maximum proton energy in L, $E_{\rm a,max}$, which in the case of
the ENDF/B library is 150 MeV, and frequently it is necessary to
extrapolate to somewhat higher energies. In {\sc penh} this
extrapolation is performed by assuming that 1) the reaction cross
section $\sigma_{\rm nr}$ and the product yields $y_{\rm b}$ are
constant, and equal to their values at $E_{\rm a,max}$, and 2)
considering the energy dependence of the average released kinetic energy
$E_{\rm kin}$, defined as the average sum of the kinetic energies of all
reaction products in the L frame. The calculation of $E_{\rm kin}$,
which we may perform by Monte Carlo simulation of a large number of
reactions, is laborious because it requires transforming the product
energies from CM to the L frame. In the simulation program we use a
simplified strategy and estimate $E_{\rm kin}$ by assuming that in the
CM frame the prompt gammas and the light products are emitted in
directions with fixed value of the polar cosine. Taking $\cos
\theta'_\gamma \simeq 0$ for the gammas and $\cos \theta'_{\rm b} \simeq
0.4$ for the light products, the value $E_{\rm kin}$ calculated in this
way approximates the exact average kinetic energy with an accuracy of
about 1\%. It is found that for proton energies $E_{\rm a}$ near the
upper end of the ENDF-6 table, $E_{\rm kin}$ can be approximated by the
formula
\beq
E_{\rm kin} =  \left[\rule{0mm}{4mm}
1 - d_1 \exp(-d_2 E_{\rm a}) \right] E_{\rm a}
\label{n.d1}\eeq
where $d_1$ and $d_2$ are parameters specific of each isotope, which are
determined from a fit to the calculated $E_{\rm kin}$ values for the
last two proton energies in the table. The parameter $d_2$ is required
to be non-negative so that $E_{\rm kin}$ generally increases with
energy tending to $E_{\rm a}$ at high energies. The simulation of
reactions induced by protons with energy $E_{\rm a} > E_{\rm a,max}$ is
performed by sampling the reaction products using the distributions for
$E_{\rm a,max}$ and multiplying their laboratory energies $E_{\rm b}$ by
the factor
\beq
F_{\rm ext}(E_{\rm a}) = \frac{1 - d_1 \exp(-d_2 E_{\rm a})}
{1 - d_1 \exp(-d_2 E_{\rm a,max})}
\, \frac{E_{\rm a}}{E_{\rm a,max}}.
\label{n.d2}\eeq
This procedure ensures that the average kinetic energy scales according to
Eq.\ \req{n.d1} and, additionally, that the reaction properties vary
continuously with the energy of the projectile proton.


\subsection{Tracking of reaction products}

In the simulation code, the kinetic energy of neutrons and of heavy
products is assumed to be deposited at the reaction site. Protons
emitted in reactions are sent to the secondary stack and subsequently
tracked. Emitted light products other than protons (deuterons, tritons,
$^3$He, and  alphas) are tracked as ``weighted equivalent'' protons as
follows.

\begin{figure}[h!] \begin{center}
\includegraphics*[width=7.5cm]{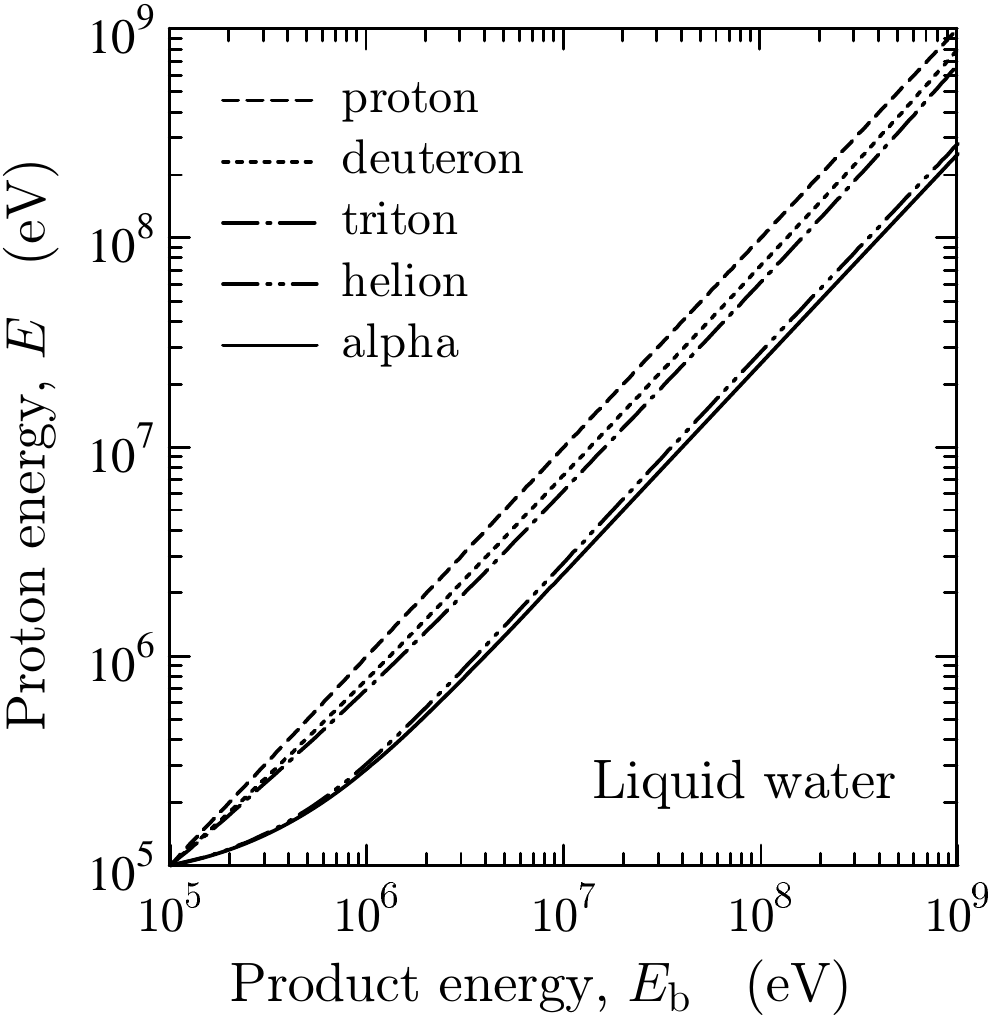}
\caption{
Energy $E$ of a proton having the same range in liquid water as a
product of energy $E_{\rm b}$ for the indicated light products.
\label{fig2}}
\end{center}\end{figure}

A light product of charge number $Z_{\rm b}$, mass $M_{\rm b}$, and kinetic
energy $E_{\rm b}$ in the L frame is replaced by a proton with initial
kinetic energy $E$ such that the proton travels the same average
distance (range) as the product. The range of charged particles is
calculated within the continuous slowing-down approximation,
\beq
R(E_{\rm b}) = \int_{E_{\rm abs}}^{E_{\rm b}} \frac{\d E}{S(E)}
\label{n.95}\eeq
where $S(E)$ is the stopping power of the product as a function of its
kinetic energy $E$, and $E_{\rm abs}$ is
the absorption energy of protons, \ie, the kinetic energy at which the
tracking of protons is discontinued and the remaining energy is assumed to
be deposited locally. The stopping power of protons is determined by the
simulation code from the DCSs of inelastic electronic excitations (see
Section 3). The stopping power of a compound material, having
${\cal N}$ molecules per unit volume and $Z$ electrons per
molecule, for light products other than the proton
is estimated from the Bethe formula
\beq
S(E) = \frac{4\pi Z_{\rm b}^2 e^4}{m_{\rm e} v^2} \, {\cal N} Z  \left[
\ln \left(\frac{2 m_{\rm e} v^2}{I} \right)
+ \ln \left( \gamma^2 \right) - \beta^2 + \frac{1}{2}
f(\gamma) - \frac{1}{2} \, \delta_{\rm F} \right],
\label{n.96}\eeq
where $m_{\rm e}$ is the electron mass, $v$ is the speed of the product, $\beta=v/c$,
$\gamma=(1-\beta^2)^{-1/2}$, $I$ is the mean
excitation energy of the material, $\delta_{\rm F}$ is the Fermi
density-effect correction (calculated from the SL GOS of the
material), and
\beq
f(\gamma) = \ln(R) + \left( \frac{m_{\rm e}}{m_{\rm b}}\,
\frac{\gamma^2-1}{\gamma} \, R \right)^2
\eeq
with
\beq
R = \left[ 1 + \left(\frac{m_{\rm e}}{M_{\rm b}} \right)^2 +
2\gamma \, \frac{m_{\rm e}}{M_{\rm b}} \right]^{-1}.
\label{n.97}\eeq
Since the formula \req{n.96} is asymptotic, \ie, it is valid only at
sufficiently high energies, it yields negative stopping powers for slow
projectiles. To prevent this anomalous behavior, for projectiles with
reduced speeds $\beta$ less than
\beq
\beta_{\rm c} = \sqrt{\frac{I}{2 m_{\rm e} c^2} \, \exp\left[ 2 - \frac{1}{2} \,
f(1) \right]}
\label{n.98}\eeq
we use the extrapolated formula
\beq
S(E) = \frac{4\pi Z_{\rm b}^2 e^4}{m_{\rm e} v^2} \, {\cal N} Z  \left[
\frac{4 \beta^2}{\beta^2 + \beta^2_{\rm c}} - \frac{1}{2} f(1)
+ \ln \left( \gamma^2 \right) - \beta^2 + \frac{1}{2}
f(\gamma) - \frac{1}{2} \, \delta_{\rm F} \right],
\label{n.99}\eeq
which smoothly matches the non-relativistic Bethe formula at
$\beta=\beta_{\rm c}$ and
yields results with the right order of magnitude for projectiles with
energies well below the range of validity of the original Bethe formula
\req{n.96}. Figure \ref{fig2} displays the energy $E$ of a proton
that has the same range as a product of energy $E_{\rm b}$ in liquid
water.

In the simulation program, a weight $w$ is attached to each particle.
Usually primary particles are given the weight $w=1$ and secondary
particles resulting from reactions inherit the weight of the parent
particle. The weight of a particle also multiplies the energy deposited
in its interactions. Most variance-reduction techniques operate by altering the
weight and number of transported particles. To ensure that the initial
kinetic energy $E_{\rm b}$ of a product with original weight $w_{\rm
b}$ is effectively transferred to the material, the ``weighted
equivalent'' proton is assigned the weight $w=w_{\rm b} (E_{\rm b}/E)$,
where $E$ is the energy of a proton having the same range
as the product. Thus, the simulation program spreads the kinetic
energy of light products within a volume around the reaction site that
has approximately the right dimensions.


\section{Sample simulation results}

The interaction models presented in the previous Sections have been
implemented in the proton transport code {\sc penh}, which works in
conjunction with the electron-photon transport code system {\sc
penelope}. The combined code performs coupled simulations of protons,
electrons/positrons, and photons in material structures consisting of
homogeneous bodies limited by quadric surfaces. The material structure
is described by using the constructive quadric geometry package {\sc
pengeom} \cite{Almansa2016, Salvat2015}. The present code differs from
the 2013 version \cite{Salvat2013} in the consideration of nuclear
effects in elastic collisions and in including proton-induced
nuclear reactions. In addition, the description of elastic collisions
has been reformulated by considering the correct relativistic equations
of the motion of the colliding particles in the CM frame.

\subsection{Dose distributions from proton beams in water}

An important application of {\sc penh} is the calculation of dose
distributions from proton beams in matter, which is of interest, \eg, in
proton therapy. To analyze the effect of the structure of the atomic
nucleus in these calculations, we present results from simulations of
200 MeV protons in water. Proton-induced nuclear reactions were
described by using the following ENDF-6 formatted files: the $^{16}$O
file from the ICRU Report 63 \cite{ICRU63} and those of $^{17}$O and
$^{18}$O from the TENDL-19 data library
(\url{https://tendl.web.psi.ch/tendl_2019/tendl2019.html}).

\begin{figure}[h!] \begin{center}
\includegraphics*[width=6.0cm]{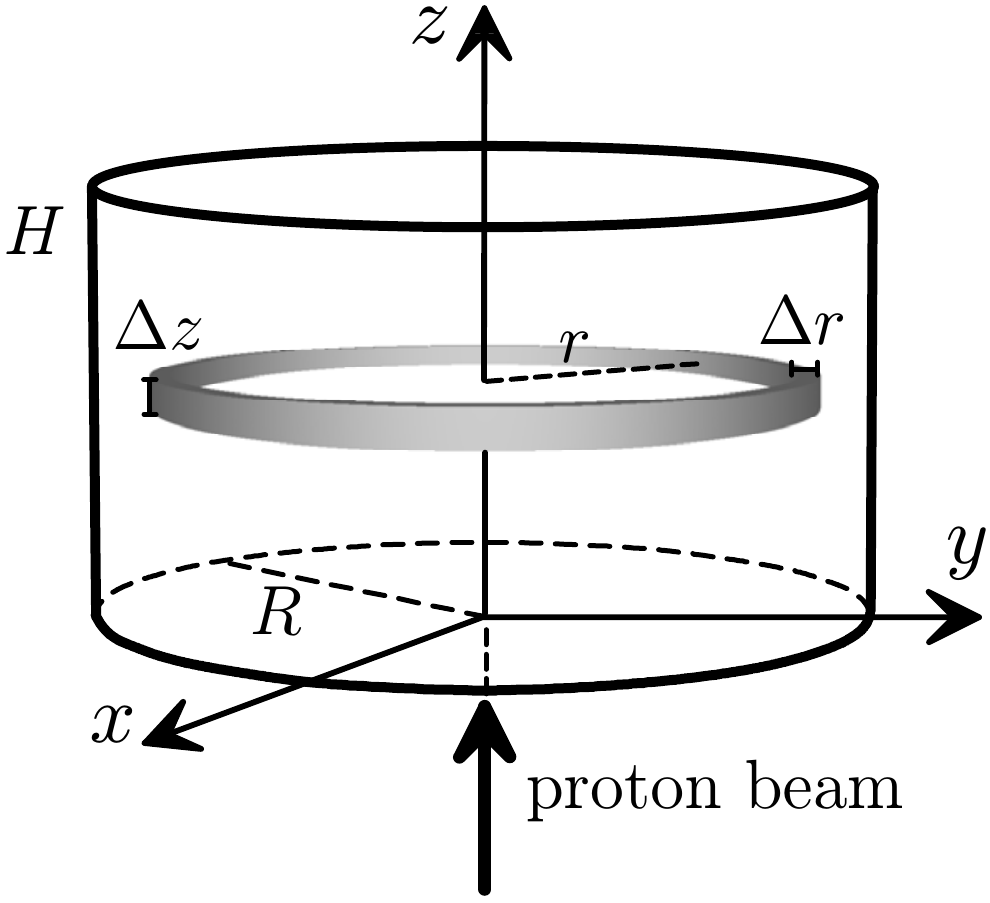}
\caption{
Schematics of the geometrical arrangement considered in the simulations.
\label{fig3}}
\end{center}\end{figure}

We consider a water cylinder of radius $R=20$ cm and height $H=40$ cm,
with the center of its base at the origin of coordinates and its
central axis coinciding with the $z$ axis, as sketched in Fig.
\ref{fig3}. A pencil beam of protons moving along the $z$ axis impinges
normally on the bottom surface of the water cylinder. Simulations were
performed by considering the following physics models: \\
1) the full model (0) described above, which accounts for the effect of
the nucleus on elastic collisions and considers proton-induced nuclear
reactions, \\
2) a simplified model (pn) in which elastic collisions are described by
considering point atomic nuclei, \ie, with the DCS given by Eq.\
\req{n.61} with $F_{\rm nuc}=1$, and nuclear reactions simulated as
in the case of the full model, and \\
3) a model (pn-nor) that assumes a point nucleus and disregards nuclear
reactions. \\
Comparison of results from simulations with these three models
illustrate the influence of nuclear effects on the transport of protons.

We have run simulations of 50 million proton histories with these three
models. The simulation speed (\ie, the number of simulated histories per
CPU second) on an Intel i7 processor at 3.4 GHz was about 300
histories/s, nearly the same for the three models. Among other
quantities, the program tallies the dose distribution $D(z,r)$ as a
function of the depth $z$ and the radial distance $r$, using volume bins
having the shape of a cylindrical ring with height $\Delta z =2$ mm and
radial thickness $\Delta r=2$ mm. Because of the cylindrical symmetry of
the arrangement, the efficiency of the simulation of the dose map is
much higher than for unsymmetrical arrangements

\begin{figure}[p!] \begin{center}
\includegraphics*[width=9.0cm]{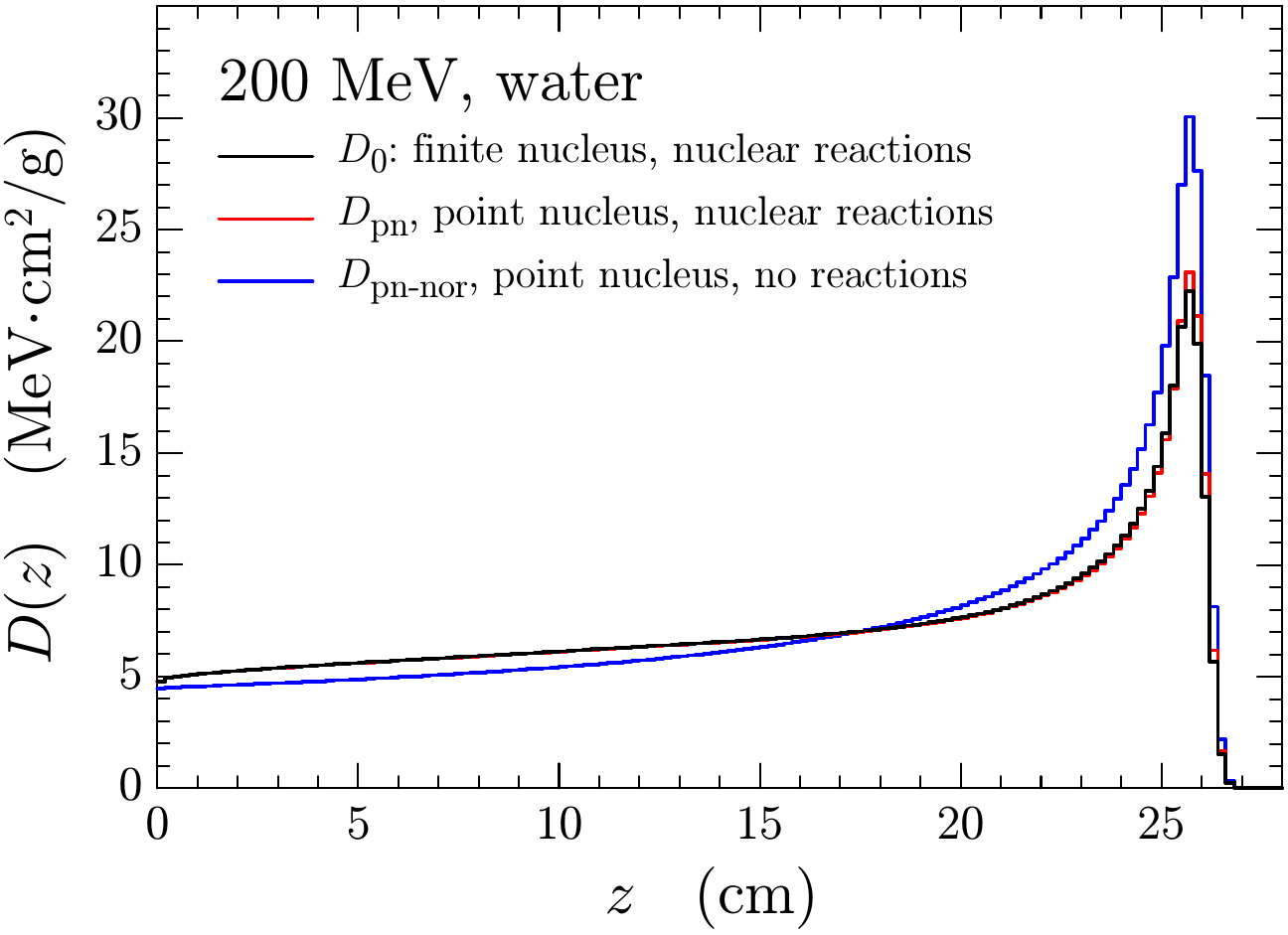} \\ [5mm]
\includegraphics*[width=9.0cm]{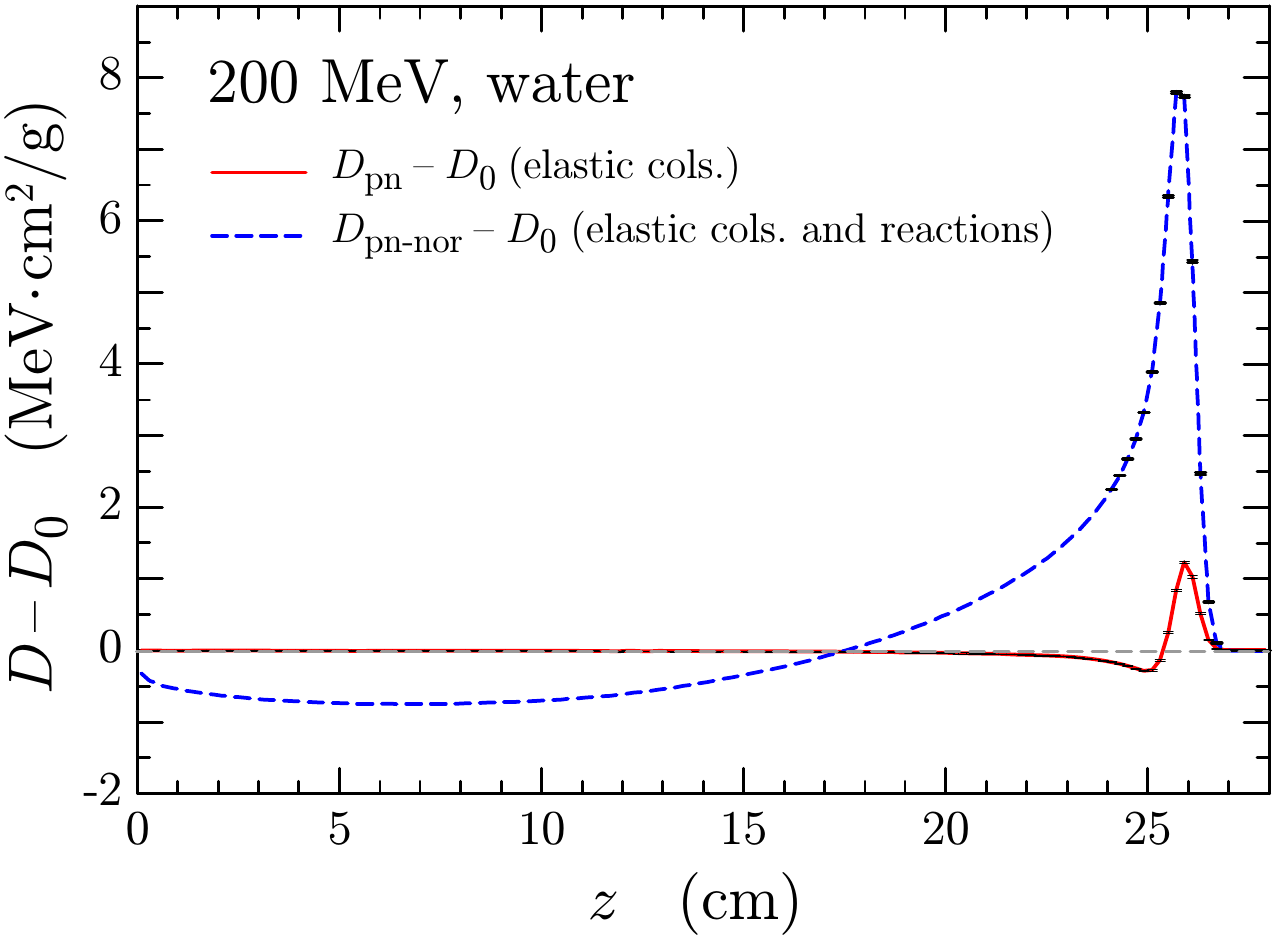}
\caption{Top: Depth-dose distributions from a 200 MeV proton beam in
water, calculated with the three physics models described in the text. Bottom:
Depth-dose differences between simulations using the full model and the
pn (point nucleus) and pn-nor (point nucleus, no nuclear reactions)
models. Statistical uncertainties, shown as error bars in the bottom
plot for $z>24$ cm, are generally less than the thicknesses of the
curves.
\label{fig4}}
\end{center}\end{figure}

Figure \ref{fig4} displays the simulated depth-dose distributions $D(z)$, \ie,
the energy deposited per unit mass thickness and per incident proton, as
a function of depth within the water cylinder, calculated as
\beq
D(z) = \int_0^R  D(z,r) \, 2\pi r\, \d r .
\eeq
The difference $D_{\rm pn}-D_0$ between the depth-doses calculated with
the pn model and with the full physics measures the effect of the
changes in the DCSs for elastic scattering that result from the finite
size and the structure of nuclei. We see that replacing the ``real''
nucleus by a point nucleus causes a relatively small variation in the
depth-dose only around the Bragg peak, which is shifted to slightly
greater depths. This result indicates that oxygen atoms with point
nuclei scatter less than atoms with real nuclei, in agreement with the
fact that $F_{\rm nuc}$ is larger than unity at intermediate angles (see
Fig.\ \ref{fig1}).  The difference $D_{\rm pn-nor} - D_0$ between the
depth-dose distributions obtained from the pn-nor model and the full
model represents the global change in $D(z)$ caused by nuclear effects.
Nuclear reactions have a strong impact on the depth-dose distributions,
they produce a significant increase of $D(z)$ at shallow depths and an
associated reduction of the primary proton flux that reaches the Bragg
peak, with a corresponding reduction of the depth-dose distribution near
that peak.

\begin{figure}[ht!] \begin{center}
\includegraphics*[width=9.0cm]{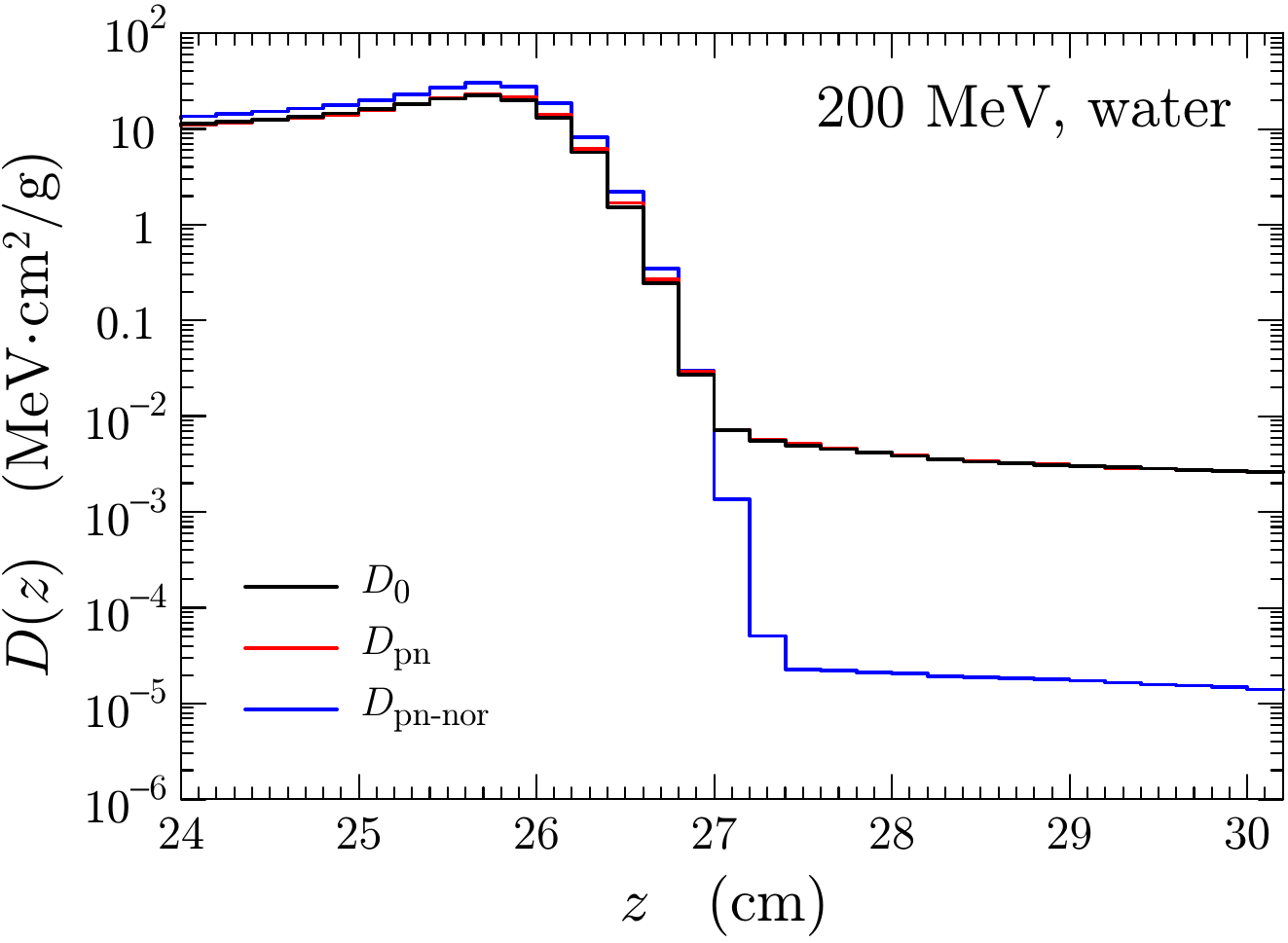}
\caption{Depth-dose distributions from a 200 MeV proton beam in water in
the Bragg peak zone and beyond, calculated with the three physics models
described in the text.
\label{fig5}}
\end{center}\end{figure}

The gammas and energetic light products released in nuclear reactions
contribute to the dose at depths beyond the Bragg peak. Figure
\ref{fig5} displays the depth-dose curves in Fig.\ \ref{fig4} with a
logarithmic scale on the vertical axis to reveal the large-$z$ tail of
$D(z)$. The pn-nor model produces a tail caused primarily by x
rays, which
extends to large depths. Nuclear reactions lead to an increase of the
large-$z$ dose by two orders of magnitude, which is due to gammas and
light reaction products. This increase in the dose may be relevant in
the planning of protontherapy treatments of tumors located in front of
sensitive organs.

\begin{figure}[p!] \begin{center}
\includegraphics*[width=11.0cm]{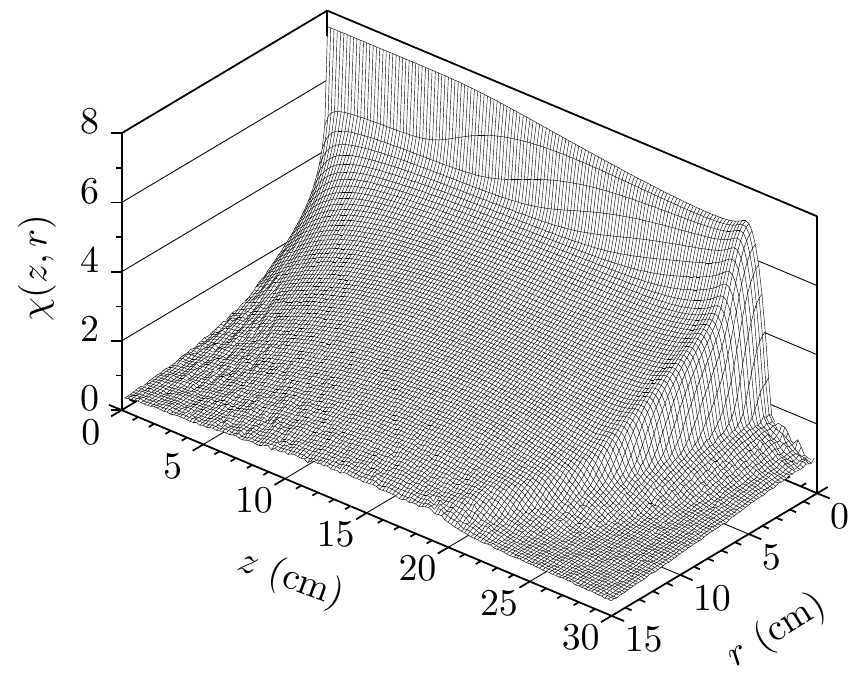}
\includegraphics*[width=11.0cm]{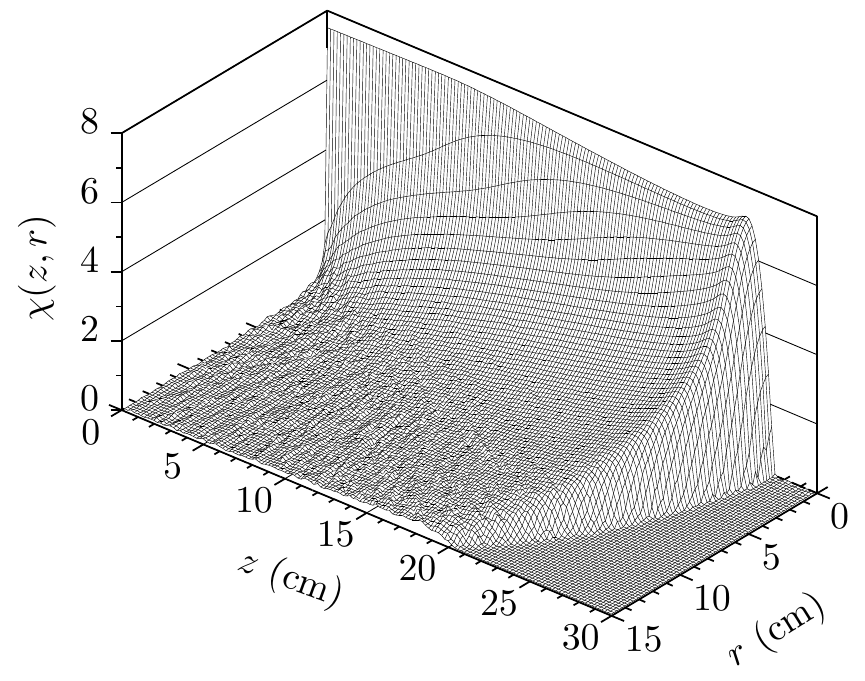}
\caption{Three-dimensional plots of the dose function $\chi(z,r)=\log_{10}
[1+D(z,r)]$, where $D(z,r)$ is the dose in units of eV/g, for a 200 MeV proton beam in
water, calculated with the full physics model (top) and with the pn-nor
model (bottom).
\label{fig6}}
\end{center}\end{figure}

Nuclear effects also alter the radial distribution of dose, the changes
become manifest when we compare dose maps obtained
from the full physics and from the pn-nor model. Figure \ref{fig6} shows
three-dimensional plots of the function
\beq
\chi(z,r) = \log_{10} \left( 1 + D(z,r) \right),
\eeq
with the dose $D(z,r)$ expressed in eV/g. Notice that $\chi \simeq D/\ln
10$ for small doses, and $\chi \simeq \log_{10} D$ when $D \gg 1$. That
is, the plots display small doses on a nearly linear scale and large
doses on a logarithmic scale. While the dose distribution calculated
with the pn-nor model falls abruptly to nearly zero behind the Bragg
peak and it decreases rapidly with the lateral distance $r$, the dose
obtained from the full physics model decreases less rapidly with
radial distance and it extends beyond the Bragg peak.
The increase of the dose beyond the volume swept by the
incident protons is caused by long-range energetic gammas and
short-range light products emitted in proton-induced nuclear reactions.


\section{Concluding comments}

The program {\sc penh} described in the present article differs from the
2013 version \cite{Salvat2013} in the consideration of nuclear effects
in elastic collisions and in including proton-induced nuclear reactions.
In addition, the description of elastic collisions has been reformulated
by considering the correct relativistic equations of the motion of the
colliding particles in the CM frame. {\sc Penh/penelope} now provides a
consistent description of electromagnetic and nuclear interactions in
proton transport simulations. The accuracy of the simulation results is
limited by the uncertainties of the adopted DCS models and of the
description of nuclear reactions given by the available ENDF-6 files,
the tracking of light products other than protons as ``weighted
equivalent'' protons, and the neglect of the production and transport of
neutrons released in nuclear reactions.  Discrepancies with proton-beam
dose measurements are to be expected at moderate and large lateral
distances from the incident beam, which are mostly attributable to the
neglected neutrons.


\section*{Acknowledgments}
\addcontentsline{toc}{section}{Acknowledgments}

We are indebted to Josep Llosa for clarifying central aspects of the
motion of two relativistic particles interacting through a central
potential in the center-of-mass reference frame.  Financial support from
the Spanish Ministerio de Ciencia, Innovaci\'{o}n y Universidades /
Agencia Estatal de Investigaci\'{o}n / European Regional Development
Fund, European Union, (projects nos.\ RTI2018-098117-B-C21 and
RTI2018-098117-B-C22) is gratefully aknowledged.



\vspace*{10mm}
\section*{References}
\bibliographystyle{elsarticle-num}
\bibliography{ACesc}

\begin{center}
\rule[-0.125mm]{1.5cm}{0.25mm}\rule[-0.3mm]{4.5cm}{.6mm}\rule[-0.125mm]{1.5cm}{0.25mm}
\end{center}

\end{document}